\documentclass[12pt]{article}
\usepackage{cite}
\usepackage{amsmath,url}
\usepackage{graphicx}
\usepackage{multirow}
\usepackage{color}
\usepackage{hyperref}
\usepackage{amssymb}
\usepackage{xspace}
\usepackage{subcaption}
\newcommand\Tstrut{\rule{0pt}{3.0ex}}         
\newcommand\Bstrut{\rule[-1.5ex]{0pt}{0pt}}   

\providecommand{\href}[2]{#2}

\newcommand\as{\alpha_{\mathrm{S}}} 
\newcommand\f[2]{\frac{#1}{#2}} 
 
\def\to{\rightarrow}

\def\pT{p_\mathrm{T}}
\def\mT{m_\mathrm{T}}

\newcommand{\GeV}{\mbox{GeV}}
\newcommand{\TeV}{\mbox{TeV}}

\newcommand{\D}{\mathrm{d}}

\newcommand{\qqbar}{q\bar{q}}
\newcommand{\qqbarprime}{q{\bar q^\prime}}

\def\refeq#1{\mbox{Eq.~\eqref{#1}}}

\def\reffi#1{\mbox{Figure~\ref{#1}}}

\def\refta#1{\mbox{Table~\ref{#1}}}

\def\refse#1{\mbox{Section~\ref{#1}}}

\def\citere#1{\mbox{Ref.~\cite{#1}}}
\def\citeres#1{\mbox{Refs.~\cite{#1}}}

\newcommand{\qT}{q_{\mathrm{T}}}
\newcommand{\qTF}{q_{\mathrm{T}}^{\mathrm{F}}}

\newcommand{\MCFM}{{\rmfamily\scshape MCFM}\xspace}
\newcommand{\Collier}{{\rmfamily\scshape Collier}\xspace}
\newcommand{\CutTools}{{\rmfamily\scshape CutTools}\xspace}
\newcommand{\OneLOop}{{\rmfamily\scshape OneLOop}\xspace}
\newcommand{\Munich}{{\rmfamily \scshape Munich}\xspace}
\newcommand{\OpenLoops}{{\rmfamily\scshape OpenLoops}\xspace}

\begin{document} 
\begin{titlepage}
\renewcommand{\thefootnote}{\fnsymbol{footnote}}
\begin{flushright}
ZU-TH 04/15\\
MITP/15-021
\end{flushright}
\vspace*{2cm}

\begin{center}
{\Large \bf $\boldsymbol{W\gamma}$ and $\boldsymbol{Z\gamma}$ production at the LHC in NNLO QCD}
\end{center}

\par \vspace{2mm}
\begin{center}
{\bf Massimiliano Grazzini$^{(a)}$\footnote{On leave of absence from INFN, Sezione di Firenze, Sesto Fiorentino, Florence, Italy.}, Stefan Kallweit$^{(b)}$} and {\bf Dirk Rathlev$^{(a)}$}
\vspace{5mm}

$^{(a)}$ Physik-Institut, Universit\"at Z\"urich, CH-8057 Z\"urich, Switzerland 

$^{(b)}$ PRISMA Cluster of Excellence, Institute of Physics,\\
Johannes Gutenberg University, D-55099 Mainz, Germany

\vspace{5mm}

\end{center}

\par \vspace{2mm}
\begin{center} {\large \bf Abstract} \end{center}
\begin{quote}
\pretolerance 10000

We consider the production of $W\gamma$ and $Z\gamma$ pairs at hadron colliders.
We report on the complete fully differential computation of radiative corrections at next-to-next-to-leading order (NNLO) in QCD perturbation theory.
The calculation includes the leptonic decay of the vector boson with the corresponding spin correlations,
off shell effects and final-state photon radiation.
We present numerical results for $pp$ collisions at 7 and 8 TeV and we compare them with available LHC data.
In the case of $Z\gamma$ production, the impact of NNLO corrections is generally moderate, ranging from 8\% to 18\%, depending on the applied cuts. In the case of $W\gamma$ production, the NNLO effects are more important, and range from 19\% to 26\%, thereby improving the agreement of the theoretical predictions with the data.
As expected, the impact of QCD radiative corrections is significantly reduced when a jet veto is applied.

\end{quote}

\vspace*{\fill}
\begin{flushleft}
April 2015

\end{flushleft}
\end{titlepage}

\setcounter{footnote}{1}
\renewcommand{\thefootnote}{\fnsymbol{footnote}}

\section{Introduction}

The discovery of a new scalar resonance in the search for the Standard Model~(SM) Higgs 
boson~\cite{Aad:2012tfa,Chatrchyan:2012ufa}
is a milestone in the LHC physics programme. The properties of this new particle
closely resemble those of the Higgs boson, but further work is needed to clarify if it is really the 
Higgs boson predicted by the SM, or something (slightly) different.
Vector-boson pair production has a prominent role in this context. It represents an irreducible background
to Higgs and new-physics searches, and, at the same time, it provides information on the form and the strength
of the vector-boson gauge couplings.
The interactions of $W$ and $Z$ bosons with photons
are particularly interesting as they test the
$WW\gamma$ and $ZZ\gamma$ couplings, which are predicted by the non-Abelian $SU(2)_L\otimes U(1)_Y$ gauge group.

Constraints on $WW\gamma$ and $ZZ\gamma$ anomalous couplings have been obtained at LEP~\cite{Schael:2013ita}. 
At hadron colliders, studies of $V\gamma$ final states have been first carried out at the 
Tevatron~\cite{Aaltonen:2011zc,Abazov:2011qp,Abazov:2011rk}, and they were used to set more stringent limits 
on anomalous couplings. The high-energy proton--proton collisions at the LHC allow us to
explore the production of $V\gamma$ ($V=W^\pm,Z$) pairs in a new energy domain. Measurements of 
$V\gamma$ final states have been carried out by ATLAS~\cite{Aad:2011tc,Aad:2012mr,Aad:2013izg,Aad:2014fha} 
and CMS~\cite{Chatrchyan:2011rr,Chatrchyan:2013fya,Chatrchyan:2013nda,Khachatryan:2015kea} using the data sets 
at centre-of-mass energy $\sqrt{s}=7$ and $8$ TeV. These measurements have been compared to the SM predictions and used to improve
the limits on anomalous couplings and on the production of possible new resonances.

When considering the $V\gamma$ final state,
besides the {\it direct} production in the hard subprocess,
the photon can also be
produced through the {\it fragmentation} of a QCD parton, and the evaluation of the ensuing contribution to 
the cross section requires the knowledge of a non-perturbative photon fragmentation function,
which typically has large uncertainties.
The fragmentation contribution is significantly suppressed by the photon isolation criteria
that are necessarily
applied in hadron-collider experiments in order to suppress the large backgrounds.
The {\it standard cone} isolation, which is 
the standard choice
in the experiments, suppresses a large fraction of the fragmentation component.
The {\it smooth cone} isolation completely suppresses the fragmentation contribution~\cite{Frixione:1998jh}, but the algorithm is difficult to be implemented experimentally.

The present status of theoretical predictions for $V\gamma$ production at hadron colliders is as follows.
The $V\gamma$ cross section is known in next-lo-leading-order~(NLO) QCD~\cite{Ohnemus:1992jn,Baur:1997kz},
and the leptonic decay of the vector boson has been included in \citere{DeFlorian:2000sg}.
In the case of $Z\gamma$ the loop-induced gluon fusion contribution, which is formally
next-to-next-to-leading order~(NNLO), has been computed in \citere{Ametller:1985di,vanderBij:1988fb},
and the leptonic decay of the $Z$ boson,
together with the gluon-induced tree-level NNLO contributions, have been added in \citere{Adamson:2002rm}.
The NLO calculation for $V\gamma$, including photon radiation from the final-state leptons, the loop-induced
gluon contribution and the photon fragmentation at LO
have been implemented into the general purpose numerical program \MCFM~\cite{Campbell:2011bn}.
Electroweak~(EW) corrections to $V\gamma$ production have been computed in \citeres{Hollik:2004tm,Accomando:2005ra}.
The full NLO EW corrections to $W\gamma$ production with leptonically decaying $W$ bosons, taking into
account all off-shell effects in the complex-mass scheme, and all effects
originating from initial-state photons, have been computed in \citere{Denner:2014bna}.
For $W\gamma$ production, the NLO computation has been matched to a parton shower according to the MiNLO 
prescription~\cite{Hamilton:2012np} in \citere{Barze:2014zba}.

In a previous letter~\cite{Grazzini:2013bna} we have presented the results of the first complete NNLO calculation 
for $Z\gamma$ production.
In the present paper, we extend this calculation to the complete class of $V\gamma$ production with leptonic decays,
namely to both $Z\gamma$ production with visible ($Z\to \ell^+\ell^-$) and invisible ($Z\to\nu\bar\nu$) $Z$-boson decays, 
and to $W\gamma$ production with the respective decays $W^+\to\nu \ell^+$ and $W^-\to \ell^-\bar\nu$. Off-shell 
effects and final-state photon radiation are consistently 
included\footnote{First results from this calculation on $W\gamma$ production have been
presented in \citere{Grazzini:2014pqa}.}.
For these production channels, we present detailed results on
fiducial cross sections and 
distributions at $\sqrt{s}=7$ and $8$ TeV, and provide comparisons to ATLAS data, where available.

The paper is organized as follows. In \refse{sec:outline} we provide some technical details of our computation and discuss 
the particular challenges in the cancellation of infrared singularities. \refse{sec:results} contains our theoretical predictions 
for all $V\gamma$ processes as well as a comparison with experimental data. We consider $Z\gamma$ production in the visible ($Z\to \ell^+\ell^-$) and 
invisible ($Z\to\nu\bar\nu$) decay channels in \refse{sec:llgam} and \refse{sec:nunugam} and $W\gamma$ production in \refse{sec:wgam}.
In \refse{sec:disc} we discuss the different impact of QCD radiative corrections in the $W\gamma$ and $Z\gamma$ processes and its physical origin. 
In \refse{sec:summary} we summarize our results and comment on the remaining uncertainties.

\section{Details of the calculation}
\label{sec:outline}

\begin{figure}[ht]
  \begin{subfigure}[b]{0.24\linewidth}
    \centering
    \includegraphics[scale=1.1,trim=1.5cm 26.0cm 2cm 1.5cm]{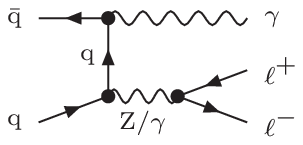} 
    \caption{Topology Ia} 
  \end{subfigure} 
  \begin{subfigure}[b]{0.24\linewidth}
    \centering
    \includegraphics[scale=1.1,trim=1.5cm 26.0cm 2cm 1.5cm]{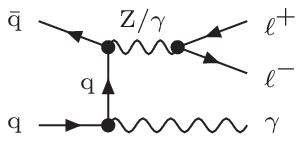} 
    \caption{Topology Ib} 
  \end{subfigure} 
  \begin{subfigure}[b]{0.24\linewidth}
    \centering
    \includegraphics[scale=1.1,trim=1.5cm 26.0cm 2cm 1.5cm]{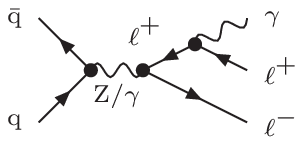} 
    \caption{Topology IIa} 
  \end{subfigure} 
  \begin{subfigure}[b]{0.24\linewidth}
    \centering
    \includegraphics[scale=1.1,trim=1.5cm 26.0cm 2cm 1.5cm]{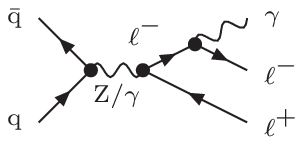} 
    \caption{Topology IIb} 
  \end{subfigure}
  \caption{Feynman diagrams contributing to $Z\gamma$ production at Born level.}
 \label{fig:zgam_diagrams} 
\end{figure}

\begin{figure}[ht]
  \begin{subfigure}[b]{0.24\linewidth}
    \centering
    \includegraphics[scale=1.1,trim=1.5cm 26.0cm 2cm 1.5cm]{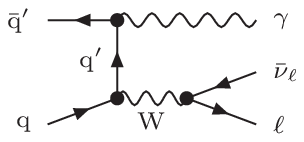} 
    \caption{Topology Ia} 
  \end{subfigure} 
  \begin{subfigure}[b]{0.24\linewidth}
    \centering
    \includegraphics[scale=1.1,trim=1.5cm 26.0cm 2cm 1.5cm]{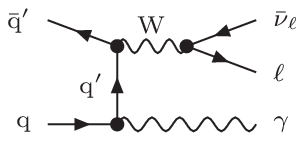} 
    \caption{Topology Ib} 
  \end{subfigure} 
  \begin{subfigure}[b]{0.24\linewidth}
    \centering
    \includegraphics[scale=1.1,trim=1.5cm 26.0cm 2cm 1.5cm]{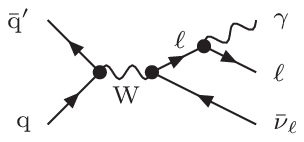} 
    \caption{Topology IIa} 
  \end{subfigure} 
  \begin{subfigure}[b]{0.24\linewidth}
    \centering
    \includegraphics[scale=1.1,trim=1.5cm 26.0cm 2cm 1.5cm]{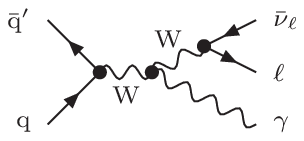} 
    \caption{Topology III} 
    \label{fig:wgam_III}
  \end{subfigure}
  \caption{Feynman diagrams contributing to $W\gamma$ production at Born level.}
 \label{fig:wgam_diagrams} 
\end{figure}

In this Section we discuss the details of our calculation. We first point out that
the notation ``$V\gamma$'' suggests
the production of an on-shell vector boson plus a photon,
followed by a factorized decay of the vector boson.
Instead, we actually compute the NNLO corrections to the processes
$pp\to \ell^+\ell^-\gamma+X$,  $pp\to \nu_\ell{\bar \nu}_\ell\gamma+X$, and $pp\to \ell\nu_\ell\gamma+X$, where, in the first case, 
the lepton pair $\ell^+\ell^-$ is produced either by a $Z$ boson or a virtual photon.
All contributions where the final-state photon is radiated off the charged leptons are consistently included (see \reffi{fig:zgam_diagrams} and \reffi{fig:wgam_diagrams}). The shortcuts ``$Z\gamma$'' and ``$W\gamma$'' are used only for convenience.

The NNLO computation requires the evaluation of tree-level scattering amplitudes with up to two additional (unresolved) partons, 
of one-loop amplitudes with up to one additional (unresolved) parton~\cite{Bern:1997sc,Campbell:2012ft}, 
and of one-loop squared and two-loop corrections to the Born subprocess
($\qqbar\to \ell^+\ell^-\gamma$ and $\qqbar\to\nu_\ell{\bar \nu_\ell}\gamma$ for $Z\gamma$, 
$\qqbarprime\to \ell\nu_\ell\gamma$ for $W\gamma$).
Furthermore, processes with charge-neutral final states receive loop-induced contributions from the gluon fusion channel 
($gg\to \ell^+\ell^-\gamma$ and $gg\to\nu_\ell{\bar \nu_\ell}\gamma$).
In our computation, all required tree-level and one-loop amplitudes are obtained from
the {\sc OpenLoops} generator~\cite{Cascioli:2011va}\footnote{The {\sc OpenLoops} one-loop generator
by F.~Cascioli, J.~Lindert, P.~Maierh{\"o}fer and S.~Pozzorini
is publicly available at \url{http://openloops.hepforge.org}.}, which
implements
a fast numerical recursion 
for the calculation of NLO scattering amplitudes within the SM. For the numerically stable evaluation of tensor integrals we rely on the {\sc Collier} library~\cite{Denner:2014gla}, which is based on the Denner--Dittmaier reduction techniques~\cite{Denner:2002ii,Denner:2005nn} and the scalar integrals of~\cite{Denner:2010tr}.

The two-loop corrections to the Drell--Yan-like Born processes, where the photon
is radiated off the final-state leptons, have been available for a long time~\cite{Matsuura:1988sm}.
The last missing ingredient,
the genuine two-loop corrections to the $V\gamma$ amplitudes,
have been presented in \citere{Gehrmann:2011ab}.

In \refse{subsec:formalism}, we give a sketch of the $\qT$ subtraction formalism, the method we use to 
combine the (separately divergent) ingredients of the NNLO calculation to obtain quantitative predictions.
The bookkeeping of all partonic subprocesses and the numerical integration of the different cross section
contributions is managed by the fully automatized 
\Munich framework~\footnote{\Munich is the abbreviation of ``MUlti-chaNnel Integrator at 
Swiss~(CH) precision''---an automated parton level NLO generator by S.~Kallweit. In preparation.}, 
which is
described in \refse{subsec:munich}. In \refse{subsec:doublevirtual}, the double-virtual and gluon fusion contributions
are discussed, and \refse{subsec:realemission} addresses the real-emission part of the NNLO cross section with
a discussion of the numerical stability issues,
namely the reliable evaluation of one-loop amplitudes 
in the real--virtual part and the numerically stable phase-space integration in the double-real part.

\subsection{The $\boldsymbol{\qT}$ subtraction formalism}
\label{subsec:formalism}
The implementation of the various scattering amplitudes in a complete NNLO calculation is
a highly non-trivial task due to the presence of infrared~(IR) singularities at
intermediate stages of the calculation, which prevents a straightforward implementation of numerical techniques.
Various methods have been proposed and used to overcome this difficulty. They are based either on extensions of the subtraction method~\cite{Frixione:1995ms,Frixione:1997np,Catani:1996jh,Catani:1996vz} at NNLO~\cite{Somogyi:2005xz,GehrmannDeRidder:2005cm,Daleo:2006xa,Currie:2013vh}, or on {\em sector decomposition}~\cite{Anastasiou:2003gr,Binoth:2000ps}. More recently, also a combination of the subtraction method with sector decomposition has been proposed~\cite{Czakon:2010td,Czakon:2011ve}.
The $\qT$ subtraction formalism~\cite{Catani:2007vq} is
an independent method to handle and cancel IR singularities at the NNLO.
In its present formulation the method applies to the
production of a colourless high-mass system $F$
in generic hadron collisions, and has been applied to the computation of
NNLO corrections to several hadronic processes~\cite{Catani:2007vq,Catani:2009sm,Ferrera:2011bk,Catani:2011qz,Grazzini:2013bna,Ferrera:2014lca,Cascioli:2014yka,Gehrmann:2014fva}.
According to the $\qT$ subtraction method~\cite{Catani:2007vq}, the $pp\to F+X$ cross section at (N)NLO 
can be written as
\begin{equation}
\label{eq:main}
\D{\sigma}^{\mathrm{F}}_{\mathrm{(N)NLO}}={\cal H}^{\mathrm{F}}_{\mathrm{(N)NLO}}\otimes \D{\sigma}^{\mathrm{F}}_{\mathrm{LO}}
+\left[ \D{\sigma}^{\mathrm{F + jet}}_{\mathrm{(N)LO}}-
\D{\sigma}^{\mathrm{CT}}_{\mathrm{(N)NLO}}\right],
\end{equation}
where $d{\sigma}^{\mathrm{F + jet}}_{\mathrm{(N)LO}}$ represents the cross section for the
production of the system $F$ plus one jet at (N)LO accuracy, and can be evaluated with
any available NLO subtraction formalism. 
The (IR subtraction) counterterm $\D{\sigma}^{\mathrm{CT}}_{\mathrm{(N)NLO}}$
is obtained from the resummation of logarithmically-enhanced
contributions to $\qT$ distributions~\cite{Bozzi:2005wk}.
The practical implementation of the contributions in the square bracket in \refeq{eq:main} is described in more detail in \refse{subsec:realemission}.

The `coefficient' ${\cal H}^{\mathrm{F}}_{\mathrm{(N)NLO}}$ encodes the loop corrections to the Born-level 
process and also compensates\footnote{More precisely, while the behavior of $\D{\sigma}^{\mathrm{CT}}_{\mathrm{(N)NLO}}$
as $\qT\to 0$ is dictated by
the singular structure of $d{\sigma}^{\mathrm{F + jet}}_{\mathrm{(N)LO}}$, its non-divergent
part in the same limit is to some extent arbitrary, and its choice determines the explicit form of ${\cal H}^{\mathrm{F}}_{\mathrm{(N)NLO}}$.}
for the subtraction of $\D{\sigma}^{\mathrm{CT}}_{\mathrm{(N)NLO}}$.
It is obtained from the (N)NLO truncation of the process-dependent perturbative function
\begin{equation}
{\cal H}^{\mathrm{F}}=1+\f{\as}{\pi}\,
{\cal H}^{\mathrm{F}(1)}+\left(\f{\as}{\pi}\right)^2
{\cal H}^{\mathrm{F}(2)}+ \dots \;\;.
\end{equation}
The NLO calculation  of $\D{\sigma}^{\mathrm{F}}$ 
requires the knowledge
of ${\cal H}^{\mathrm{F}(1)}$, and the NNLO calculation also requires ${\cal H}^{\mathrm{F}(2)}$.
The general structure of ${\cal H}^{\mathrm{F}(1)}$
is known~\cite{deFlorian:2001zd}: 
${\cal H}^{\mathrm{F}(1)}$ is obtained from the process-dependent scattering
amplitudes by using a process-independent relation.
Exploiting the explicit results of ${\cal H}^{\mathrm{F}(2)}$ for Higgs
\cite{Catani:2011kr} and vector boson~\cite{Catani:2012qa} 
production,
the process-independent relation of 
\citere{deFlorian:2001zd} has been extended to the calculation of the NNLO coefficient 
${\cal H}^{\mathrm{F}(2)}$~\cite{Catani:2013tia}.
These results have been confirmed through an independent calculation in
the framework of Soft-Collinear Effective Theory (SCET)~\cite{Gehrmann:2012ze,Gehrmann:2014yya}.
The counterterm $\D{\sigma}^{\mathrm{CT}}_{\mathrm{(N)NLO}}$ 
only depends on ${\cal H}^{\mathrm{F}}_{\mathrm{(N)LO}}$, i.e.\ for a NNLO computation, it requires only 
${\cal H}^{\mathrm{F}(1)}$ as input, which can be derived from the one-loop amplitudes to the 
involved Born subprocesses.

As can be seen from the discussion above, a significant part of the ingredients needed to perform 
the NNLO computation are actually NLO-like in nature, allowing us to build the 
implementation of our NNLO calculation upon existing NLO tools. We have based this calculation on the \Munich framework, which 
is briefly discussed in the following. Necessary extensions of this framework are addressed as well in
the rest of this Section.

\subsection{Organization of the calculation within the {\sc Munich} framework}
\label{subsec:munich}
\Munich is a fully automatized framework for the computation of fixed-order 
cross sections for arbitrary SM processes up to NLO accuracy, written in C++.
After the hadronic process has been specified, 
\Munich takes care of the bookkeeping of all partonic subprocesses that need to
be taken into account.
It automatically generates adequate phase-space parametrizations for each partonic subprocess 
by exploiting the resonance structure of the underlying (squared) tree-level Feynman diagrams.
These parametrizations are combined using a multi-channel approach 
to simultaneously flatten the, in general complicated, resonance structure of the 
amplitudes and thus guarantee a reasonable convergence of the numerical integration.
Several improvements like an adaptive weight-optimization procedure are implemented as well.

\Munich was originally developed for the NLO calculations of~\cite{Denner:2010jp,Denner:2012yc,Denner:2012dz}, where only massless 
colour-charged particles were involved. To account for the mediation of IR singularities between
the different phase spaces of virtual and real corrections, the Catani--Seymour dipole 
subtraction method~\cite{Catani:1996jh,Catani:1996vz} was implemented.
In \citere{Cascioli:2013wga}, the framework was extended to massive QCD according to \citere{Catani:2002hc}, and 
in \citere{Kallweit:2014xda} the extension to complete SM calculations, i.e.\ including also
EW corrections at NLO, was presented.
All dipole terms necessary to numerically cancel the soft and collinear 
divergences of the real corrections are constructed at runtime from spin- and colour-correlated matrix elements 
of the underlying partonic Born subprocesses.
Moreover, additional phase-space parametrizations based on the reduced dipole kinematics are generated 
and included in the multi-channel, supplementing the parametrizations based on real-emission kinematics.
Analogously, the analytically integrated dipoles, the so-called $K+P$ terms and the $I$-operator, 
which compensate for the dipole terms subtracted on the real-emission side, 
are automatically constructed at runtime from colour-correlated matrix elements of the Born subprocesses.
For the evaluation of all involved matrix elements up to the one-loop order, \Munich provides 
an automatic interface to amplitudes generated by the \OpenLoops generator~\cite{Cascioli:2011va}.

The guiding principle in our NNLO implementation is to keep the additionally introduced process dependence 
to the bare minimum, essentially limiting it to the two-loop amplitudes entering $\mathcal{H}^{\mathrm{F}}$ in \refeq{eq:main}. 
All other process dependent information entering the various pieces in \refeq{eq:main} have been expressed 
in terms of NLO quantities available inside \Munich via its interface to \OpenLoops.

\subsection{Double-virtual and gluon fusion contribution}
\label{subsec:doublevirtual}
In our implementation of the $\qT$ subtraction formalism, the only component that needs to be provided on a process-by-process 
basis are the two-loop---and one-loop-squared\footnote{
The \OpenLoops generator also provides the non-finite one-loop-squared matrix elements in terms of 
the corresponding coefficients of the Laurent series. These results are, however, only used as a numerical
check in this calculation.
}---amplitudes to the Born processes, which enter the coefficient 
${\cal H}^{\mathrm{F}}_{\mathrm{(N)NLO}}$ in \refeq{eq:main}.
\citere{Gehrmann:2011ab} provides the 
analytical expressions of the resonant helicity amplitudes for $V\gamma$ production 
(e.g.\ $\qqbar\to Z\gamma \to \ell^+\ell^-\gamma$) in terms of two-dimensional harmonic polylogarithms up to weight 4. 
The non-resonant (final-state radiation) contribution (e.g.\ $\qqbar\to \ell^+\ell^-\to \ell^+\ell^-\gamma)$ 
is described by the two-loop quark form factor from \citere{Matsuura:1988sm}. 
We have implemented the helicity amplitudes directly into a C++ library. For the numerical evaluation of the 
harmonic polylogarithms we use the \verb+tdhpl+ library~\cite{Gehrmann:2001jv}. Our implementation 
allows for several thousand amplitude evaluations per minute, rendering the time spent
on computing the two-loop contribution negligible compared to the time needed for the real emission contribution.

For the processes with an electrically neutral final state at Born level, 
i.e.\ $pp\to \ell^+\ell^-\gamma$ and $pp\to \nu_\ell\overline{\nu}_\ell\gamma$, 
the loop-induced gluon fusion processes
$gg \to \ell^+\ell^-\gamma$ and $gg \to \nu_\ell\overline{\nu}_\ell\gamma$
represent a separately IR-finite and gauge-invariant part of the cross section. 
Though of $\mathcal{O}\left(\as^2\right)$, and thus formally of NNLO, the gluon fusion contribution 
is often included already at NLO (see for example \MCFM~\cite{Campbell:2011bn}.) 
The reason for that is the assumption that the large gluon luminosities at the LHC could compensate for the 
additional $\as$ suppression, and hence the contribution could numerically become as important as the 
NLO corrections themselves. 
As we will see in \refse{sec:llgam} and \refse{sec:nunugam}, this is not the case in $Z\gamma$ production. 
The gluon fusion contribution is small even when compared to the other NNLO corrections. 
In our computation, these (finite) one-loop-squared gluon fusion amplitudes are evaluated using \OpenLoops.

\subsection{Real-emission and counterterm contribution}
\label{subsec:realemission}
All NNLO contributions with vanishing total transverse momentum $\qT$ of the final state system $F$
are collected in the coefficient ${\cal H}^{\mathrm{F}}_{\mathrm{NNLO}}$, which is discussed in the previous Section.
The remaining part of the NNLO cross section, namely the difference in the square bracket in \refeq{eq:main},
is formally finite in the limit $\qT\to 0$, but the terms separately exhibit logarithmic divergences in this limit.
The counterterm $\D{\sigma}^{\mathrm{CT}}_{\mathrm{(N)NLO}}$ is integrated over the $n$-particle Born phase-space\footnote{The integration over $\qT$ only acts on an explicit $\qT$ dependence of the phase-space weight.}, while 
the term $\D{\sigma}^{\mathrm{F + jet}}_{\mathrm{NLO}}$ in \refeq{eq:main}, called the real-emission contribution
in the following, involves two different phase spaces with one and two additional QCD partons, respectively, and thus 
needs to be integrated separately over the respective \mbox{$(n+1)$}- and \mbox{$(n+2)$}-particle phase spaces. To achieve 
a numerical cancellation of the singularity, a technical cut on $\qTF$ needs to be introduced to render both terms separately finite. 
In this sense, the $\qT$ subtraction method works very similarly to a phase-space slicing method at NLO.
In practice, a technical cut, $r_{\mathrm{cut}}$, on the dimensionless quantity $r\equiv \qT/M$, 
where $M$ denotes the invariant mass of the final-state system $F$, 
turns out to be a more convenient choice than a cut on $\qT$ itself.

By construction, the counterterm $\D{\sigma}^{\mathrm{CT}}_{\mathrm{(N)NLO}}$ cancels all 
$\qT$-divergent (logarithmic) terms from the real-emission contributions, 
implying that the $r_{\mathrm{cut}}$ dependence of their difference does not only vanish in the limit $r_{\mathrm{cut}}\to0$, but
should become arbitrarily small for sufficiently small values of $r_{\mathrm{cut}}$.
In practice, however, as both the counterterm and the real-emission contribution grow arbitrarily large for $r_{\mathrm{cut}}\to0$, 
the statistical accuracy of the Monte Carlo integration degrades, preventing one from pushing $r_{\mathrm{cut}}$ arbitrarily low.
In general, the absence of any strong residual $r_{\mathrm{cut}}$ dependence in the difference between the real contribution 
$\D{\sigma}^{V\gamma+{\rm jet}}_{\mathrm{(N)LO}}$ and the counterterm $\D{\sigma}^{\mathrm{CT}}_{\mathrm{(N)NLO}}$ 
provides a strong check on the correctness of the computation since any (significant) mismatch between the 
contributions would result in a divergent cross section in the limit $r_{\mathrm{cut}}\to0$.
To monitor the $r_{\mathrm{cut}}$ dependence without the need of several CPU-intensive runs,
our implementation allows for simultaneous cross section evaluations at arbitrary $r_{\mathrm{cut}}$ values.

Typical values of $r_{\mathrm{cut}}$ are of the order of $10^{-3}$, requiring a numerically stable 
evaluation of $\D{\sigma}^{V\gamma+{\rm jet}}_{\mathrm{NLO}}$ down to the IR-divergent region where 
the transverse-momentum of the $V\gamma$ system reaches values of about $0.1\,\GeV$.
While the IR-enhanced phase-phase region with a very-low-$\qT$ jet is excluded in standard NLO calculations, 
it needs to be resolved at very high precision in a NNLO computation. This issue
obviously challenges a dedicated NLO program like \Munich. We have slightly modified
the phase-space generation to achieve the required precision and reliability of 
our NNLO results within the \Munich framework.

The real-emission contribution to the NNLO cross section is, apart from the presence of a very-low-$\qT$ jet discussed above,
a usual NLO calculation, and can thus be treated with the Catani--Seymour dipole subtraction method implemented within \Munich.
Consequently, it can be split into a real--virtual~(RV), a real--collinear~(RC) and a 
double-real~(RR) contribution,
\begin{align}
\D{\sigma}^{V\gamma+{\rm jet}}_{\mathrm{NLO}} = \D{\sigma}^{V\gamma}_{\mathrm{RV}} + \D{\sigma}^{V\gamma}_{\mathrm{RC}} + \D{\sigma}^{V\gamma}_{\mathrm{RR}}.
\end{align}
Here, the mediation of all IR divergences (given a finite $r_{\mathrm{cut}}$ value) by dipole subtraction is implicitly understood, 
i.e.\ each contribution on the right-hand side is finite and can be numerically integrated over the respective phase space.

Whereas the real-collinear subcontribution does not exhibit any peculiar issues,
the other two subcontributions on the right-hand side pose different challenges 
when integrated into the deep IR ($\qT\gtrsim 0$) region. 
The real--virtual contribution requires the evaluation of one-loop matrix elements
to $V\gamma+{\rm jet}$ production, which has become a standard task for
automatic one-loop tools.
However, as we evaluate the matrix elements in the deep IR region, i.e.\ 
far away from the phase-space region 
that is relevant for $2\to 3$ hard scattering processes at NLO,
numerical instabilities in the amplitude evaluation might be a source of concern. To address this issue, 
\OpenLoops implements a fully automated system that monitors the numerical accuracy
of loop amplitudes and cures possible instabilities at runtime.
This system exploits the fact that, for the reduction to 
scalar integrals, OpenLoops allows one to flexibly switch from tensor-reduction~\cite{Denner:2002ii,Denner:2005nn}
to OPP reduction~\cite{Ossola:2006us} algorithms.

To perform the reduction from tensor to scalar integrals, 
we make use of the \Collier library~\cite{Denner:2014gla} that consists of
two independent implementations of the Denner--Dittmaier reduction algorithm~\cite{Denner:2002ii,Denner:2005nn}. 
The presence of potential residual instabilities
is tested by comparing the two implementations of tensor reduction. 
In presence of instabilities that exceed one permille of the Born amplitude,
the one-loop amplitude is reevaluated using the \CutTools~\cite{Ossola:2007ax} implementation of 
OPP reduction in quadruple precision. In this case \OneLOop~\cite{vanHameren:2010cp} is used for the evaluation
of scalar integrals. Finally, the accuracy of the result in 
quadruple precision is verified by a consistency check based on the
rescaling of all dimensionful variables.
For all processes considered in this paper we find that the Denner--Dittmaier reduction 
works fast and reliably for more than 99\% of the phase space points.
This allows one to restrict the usage of quadruple precision to a 
tiny fraction of points with a minor impact on the total runtime of the calculation.

The double-real contribution, on the other hand, only involves tree-level amplitudes, but contains an additional 
unresolved parton in the final state. Moreover, it involves several dipole terms located on different 
$(n+1)$-particle phace spaces. Like any other phase-space cut, the restriction $r>r_{\mathrm{cut}}$ has to be applied on the
respective phase space and can thus lead to miscancellation issues, which are well-known from 
NLO calculations, but might be amplified here when happening in the deep IR region. These complications 
render the numerical phase-space integration in the deep IR region more 
challenging compared to the integration of the single-emission phase space.
\Munich already provides a state-of-the-art multi-channel phase-space integrator, 
which greatly speeds up computations compared to a classical VEGAS integration procedure,
and guarantees a reasonably stable and reliable convergence behaviour.
In the deep IR regions, however, even an advanced multi-channel approach might fail to capture all 
relevant contributions. We have thus extended the \Munich integrator by an additional 
VEGAS-like importance sampling on top of the multi-channel parametrization. 
This hybrid approach results in an extremely efficient and reliable 
phase-space integration.

The numerical information on the $r_{\mathrm{cut}}$ dependence of the cross section is used to perform an
extrapolation to $r_{\mathrm{cut}}=0$, which can in turn be used to quantify the uncertainty due to the finite $r_{\mathrm{cut}}$ value.
This estimated uncertainty is combined with the usual statistical
integration error to provide an overall estimate of the numerical precision of our NNLO prediction. For the processes considered in this paper, the extrapolation at $r_{\mathrm{cut}}\to 0$ turns out to be non-trivial, due to the
interplay with the photon isolation. Nonetheless, the procedure allows us
to control integrated NNLO cross sections to better than $1\%$.

\section{Results}
\label{sec:results}
\subsection{Setup}
\label{setup}
For the electroweak couplings we use the so-called $G_\mu$ scheme,
where the input parameters are $G_F$, $m_W$, $m_Z$. In particular we 
use the values
$G_F = 1.16639\times 10^{-5}$~GeV$^{-2}$, $m_W=80.399$ GeV,
$m_Z = 91.1876$~GeV, $\Gamma_Z=2.4952$~GeV and $\Gamma_W=2.1054$~GeV. We set the CKM matrix to unity.
We use the MMHT 2014~\cite{Harland-Lang:2014zoa} sets of parton distribution functions (PDFs), with
densities and $\as$ evaluated at each corresponding order
(i.e., we use $(n+1)$-loop $\as$ at N$^n$LO, with $n=0,1,2$),
and we consider $N_f=5$ massless quarks/antiquarks and gluons in 
the initial state.

The default
renormalization ($\mu_R$) and factorization ($\mu_F$) scales are set to
$\mu_R=\mu_F=\mu_0\equiv\sqrt{m_V^2+(\pT^{\gamma})^2}$. 
An estimate of missing higher-order contributions is obtained by performing
$\mu_F$ and $\mu_R$ variations by a factor of two around the central value.
We find substantial cancellations between
renormalization and factorization scale variations in some of the calculations we are going to present
if the restriction
$\mu_R=\mu_F$ is imposed. These cancellations are assumed to be purely accidental.
To accomodate this well-known feature, we consider also antipodal variations of the two 
scales~\cite{Campbell:2011bn}, i.e.\ setting $\mu_R=\xi\mu_0$, 
$\mu_F=\mu_0/\xi$ and varying $\xi$ between $\frac{1}{2}$ and $2$.
In summary, we estimate scale uncertainties by varying $\mu_F$ and $\mu_R$ simultaneously 
and independently in the range $0.5\mu_0$ and $2\mu_0$ with no constraint on the ratio $\mu_F/\mu_R$.

The present formulation of the $\qT$ subtraction formalism~\cite{Catani:2007vq}
is limited to the production of colourless systems $F$ and, hence, it does not
allow us to deal with the 
parton fragmentation subprocesses.
Therefore, we consider only direct photons, and 
we rely on the smooth cone isolation criterion~\cite{Frixione:1998jh}. Considering a cone 
of radius $r=\sqrt{(\Delta \eta)^2+(\Delta \phi)^2}$ around the photon, 
we require that the total amount of hadronic (partonic) transverse energy $E_T$ 
inside the cone is smaller than $E_{T}^{\rm max}(r)$,
\begin{equation}
E_{T}^{\rm max}(r) \equiv  \epsilon_\gamma \,\pT^\gamma \left(\frac{1-\cos r}{1- \cos R}\right)^n \, ,
\label{eq:frixione}
\end{equation}
where $\pT^\gamma$ is the photon transverse momentum; the isolation criterion
$E_T < E_{T}^{\rm max}(r)$ has to be fulfilled for all cones with $r\leq R$.
All results presented in this paper are obtained
with $\epsilon_\gamma=0.5$, $n=1$ and $R=0.4$.

\subsection{Comparison to experimental data}

The smooth cone isolation prescription adopted in our calculation is not yet implemented
in experimental analyses.
Measurements are usually performed by using a fixed cone isolation prescription (which corresponds to \refeq{eq:frixione} with $n=0$),  and thus
our isolation prescription does not exactly correspond to what is done in the experiment.
However, the parameters $\epsilon_\gamma$ and $R$ needed 
to specify the smooth cone have natural counterparts in the definition of the fixed cone, while the precise choice of the 
smoothing function (in our case parametrized by $n$) does only have a mild impact on the final result. Furthermore, recent studies carried out in diphoton production~\cite{Butterworth:2014efa}
show that for sufficiently tight isolation parameters, smooth and hard 
cone isolation yield very similar results. For the processes in this paper, we verified at NLO that the difference between using 
smooth and hard cone isolation is at the $1-2\%$ level\footnote{Obviously, the agreement also significantly depends on the fragmentation 
function used when employing the hard cone isolation, which typically has large uncertainties.},
i.e.\ well below the current experimental uncertainties and still smaller than
the remaining theoretical uncertainties.
We can thus safely compare our theoretical predictions with experimental data.

Since the first results of our work have appeared~\cite{Grazzini:2013bna,Grazzini:2014pqa},
we have provided numerical predictions
for $Z\gamma$ production to the CMS collaboration~\cite{Khachatryan:2015kea}, and for $W\gamma$ production 
to the ATLAS collaboration~\cite{ATLAS:2014aga}, as a background in the $H\to WW$ analysis.
These predictions were obtained by using the experimental cuts adopted in the corresponding analyses.
In the present paper, we limit ourselves to compare our predictions to the ATLAS results for $W\gamma$ and $Z\gamma$ at 7 TeV~\cite{Aad:2013izg}.

\subsection{\texorpdfstring{$pp\to \ell^+\ell^-\gamma$}{llgamma production}}
\label{sec:llgam}

In our calculation of $pp\to \ell^+\ell^-\gamma$ at
$\sqrt{s}=7$ and $8$ TeV we adopt the selection cuts used by
the ATLAS collaboration~\cite{Aad:2013izg}, summarized in \refta{tab:zll_cuts}.
We require the photon to have a transverse momentum of $\pT^\gamma>15\,\GeV$ (soft $\pT^\gamma$ cut) or $\pT^\gamma>40\,\GeV$ (hard $\pT^\gamma$ cut) and pseudorapidity $|\eta^\gamma|<2.37$. 
Each of the charged leptons is required to have $\pT^\ell>25$ GeV and $|\eta^\ell|<2.47$, and the invariant mass of the lepton pair 
must fulfill $m_{\ell^+\ell^-}>40$ GeV.
We require the separation in rapidity and azimuth $\Delta R$ between the leptons and the photon to be 
$\Delta R(\ell,\gamma)>0.7$.
Jets are reconstructed with the anti-$k_T$ algorithm~\cite{Cacciari:2008gp} with radius parameter $D=0.4$. 
A jet must have $\pT^{\rm jet}>30$ GeV 
and $|\eta^{\rm jet}|<4.4$. We require the separation $\Delta R$ between 
the leptons (photon) and the jets to be $\Delta R(\ell/\gamma,{\rm jet})>0.3$.
At $\sqrt{s}=8\,\TeV$, the jet definition is slightly adjusted by using $|\eta^{\rm jet}|<4.5$ instead of $|\eta^{\rm jet}|<4.4$,
adapting to the ATLAS Run II standard.
With respect to resolved jets in the final states, we will consider both the 
inclusive ($N_{\rm jet}\geq0$) and the exclusive ($N_{\rm jet}=0$) case.

\begin{table}[tp]
\begin{center}
\begin{tabular}{lcc}
& \hspace*{3em}$\sqrt{s}=7\,\TeV$\hspace*{3em} & \hspace*{3em}$\sqrt{s}=8\,\TeV$\hspace*{3em}%
\Bstrut
\\
\hline

\Tstrut
Leptons
& \multicolumn{2}{c}{$\pT^\ell>25\,\GeV$}%
\Bstrut
\\
& \multicolumn{2}{c}{$|\eta|<2.47$}%
\Bstrut
\\
\hline

\Tstrut
Photon
& \multicolumn{2}{c}{$\pT^\gamma>15\,\GeV$ (soft $\pT^\gamma$ cut) or $\pT^\gamma>40\,\GeV$ (hard $\pT^\gamma$ cut)}%
\Bstrut
\\
& \multicolumn{2}{c}{$|\eta^\gamma|<2.37$}%
\Bstrut
\\
& \multicolumn{2}{c}{Frixione isolation with $\varepsilon_\gamma=0.5$, $R=0.4$, $n=1$}%
\Bstrut
\\
\hline

\Tstrut
Jets 
& \multicolumn{2}{c}{anti-$k_\mathrm{T}$ algorithm with $D=0.4$}%
\Bstrut
\\
& \multicolumn{2}{c}{$\pT^{\rm jet}>30\,\GeV$}%
\Bstrut
\\
& $|\eta^{\rm jet}|<4.4$ & $|\eta^{\rm jet}|<4.5$%
\Bstrut
\\
& \multicolumn{2}{c}{$N_{\rm jet}\geq 0$ (inclusive) or $N_{\rm jet}=0$ (exclusive)}%
\Bstrut
\\
\hline
\Tstrut
Separation
& \multicolumn{2}{c}{$m_{\ell^+\ell^-}>40\,\GeV$}%
\Bstrut
\\
& \multicolumn{2}{c}{$\Delta R(\ell,\gamma)>0.7$}%
\Bstrut
\\
& \multicolumn{2}{c}{$\Delta R(\ell/\gamma,{\rm jet})>0.3$}%
\Bstrut
\\
\hline
\end{tabular}

\end{center}
\caption{Event selection criteria used in the $pp\to \ell^+\ell^-\gamma$ analysis.}
\label{tab:zll_cuts}
\end{table}

The predicted cross sections with the soft $\pT^\gamma$ cut, including the theoretical uncertainties 
from scale variations obtained as described at the beginning of \refse{setup}, can be found in 
\refta{tab:zll_results}. When going from NLO to NNLO the cross section increases by 8\% (3\%) 
in the inclusive (exclusive) case, respectively.
The fiducial cross sections measured by ATLAS at 7 TeV~\cite{Aad:2013izg} are also reported in \refta{tab:zll_results}. 
Both the NLO and NNLO predictions are in agreement with the experimental result,
and the NNLO corrections improve the agreement, especially in the inclusive case.

The reduced impact of QCD radiative corrections when going from the inclusive ($N_{\rm jet}\geq 0$) to the
exclusive ($N_{\rm jet}=0$) case is a well known feature in perturbative QCD calculations~\cite{Catani:2001cr}.
A stringent veto on the radiation recoiling against the $Z\gamma$ system tends to unbalance the cancellation
between positive real and negative virtual contributions, possibly leading to large logarithmic terms.
The resummation of these logarithmic contributions has been the subject of intense theoretical studies~\cite{Banfi:2012jm,Becher:2012qa,Stewart:2013faa},
especially in the important case of Higgs boson production.
The reduced impact of radiative effects in the presence of a jet veto is often accompanied
by a reduction of scale uncertainties.
In the present case, since we are considering a process initiated by quark-antiquark scattering, the
impact of radiative corrections is smaller than in Higgs boson production. However, a reduction of scale uncertainties from the $N_{\rm jet}\geq 0$ to the $N_{\rm jet}= 0$ case is already visible in \refta{tab:zll_results},
and may signal the need of more sophisticated (conservative) methods to estimate perturbative uncertainties~\cite{Stewart:2011cf,Banfi:2012jm}.

\begin{table}[tp]
\begin{center}
\begin{tabular}{c c c c c c}
$\sqrt{s}$ [TeV]
&
& $\sigma_{\mathrm{LO}}$ [pb] 
& $\sigma_{\mathrm{NLO}}$ [pb] 
& $\sigma_{\mathrm{NNLO}}$ [pb]
& $\sigma_{\mathrm{ATLAS}}$ [pb]
\Bstrut
\\ 
\hline
\Tstrut
\multirow{3}{*}{7}
& $N_{\rm jet}\geq 0$ 
& \multirow{3}{*}{$0.8149^{+8.0\%}_{-9.3\%}$} 
& $1.222^{+4.2\%}_{-5.3\%}$
& $1.320^{+1.3\%}_{-2.3\%}$
& $1.31 \begin{array}{l}\\[-3ex]\scriptstyle\pm 0.02~{\rm (stat)}\\[-1ex]\scriptstyle\pm 0.11~{\rm (syst)}\\[-1ex] \scriptstyle\pm 0.05~{\rm (lumi)}\end{array}$
\Bstrut
\\
& $N_{\rm jet} = 0$ 
&  
& $1.031^{+2.7\%}_{-4.3\%}$
& $1.059^{+0.7\%}_{-1.4\%}$
& $1.05 \begin{array}{l}\\[-3ex]\scriptstyle\pm 0.02~{\rm (stat)}\\[-1ex] \scriptstyle\pm 0.10~{\rm (syst)}\\[-1ex] \scriptstyle \pm 0.04~{\rm (lumi)}\end{array}$
\Bstrut
\\
\hline
\Tstrut
\multirow{2}{*}{8}
& $N_{\rm jet}\geq 0$ 
& \multirow{2}{*}{$0.9244^{+9.0\%}_{-10.2\%}$}
& $1.387^{+4.3\%}_{-5.7\%}$
& $1.504^{+1.3\%}_{-2.5\%}$
& 
\Bstrut
\\
& $N_{\rm jet} = 0$ 
&  
& $1.157^{+2.6\%}_{-4.5\%}$
& $1.188^{+0.8\%}_{-1.5\%}$ 
& 
\Bstrut
\\
\hline
\end{tabular}

\end{center}
\caption{$pp\to \ell^+\ell^-\gamma$ cross sections with the soft $\pT^\gamma$ cut ($\pT^\gamma>15\,\GeV$). 
Scale uncertainties are obtained from independent variations of $\mu_R$ and $\mu_F$
around the central scale $\mu_0$, as described in \refse{setup}. 
The numerical uncertainty of the NNLO prediction from statistical error and finite $r_{\mathrm{cut}}$ are conservatively estimated to be about $0.3\%$.
The last column provides the corresponding results by ATLAS.}
\label{tab:zll_results}
\end{table}

\begin{table}[tp]
\begin{center}
\begin{tabular}{c c c c c c}
$\sqrt{s}$ [TeV]
&
& $\sigma_{\mathrm{LO}}$ [fb] 
& $\sigma_{\mathrm{NLO}}$ [fb] 
& $\sigma_{\mathrm{NNLO}}$ [fb]
\Bstrut
\\ 
\hline
\Tstrut
7
& $N_{\rm jet}\geq 0$ 
& $73.61^{+3.4\%}_{-4.5\%}$
& $132.0^{+4.2\%}_{-4.0\%}$
& $154.3^{+3.1\%}_{-2.8\%}$
\Bstrut
\\
\hline
\Tstrut
8
& $N_{\rm jet}\geq 0$ 
& $84.09^{+4.3\%}_{-5.5\%}$
& $153.1^{+4.6\%}_{-4.5\%}$
& $180.1^{+3.1\%}_{-3.0\%}$
\Bstrut
\\
\hline
\end{tabular}

\end{center}
\caption{$pp\to \ell^+\ell^-\gamma$ cross sections with the hard $\pT^\gamma$ cut ($\pT^\gamma>40\,\GeV$). 
Scale uncertainties are computed as in \refta{tab:zll_results}.
The numerical uncertainty of the NNLO prediction from statistical error and finite $r_{\mathrm{cut}}$ is conservatively estimated to be about $0.6\%$.
}
\label{tab:zll_results_hard}
\end{table}

\begin{figure}[tp]
  \centering
  \includegraphics[width=0.5\columnwidth]{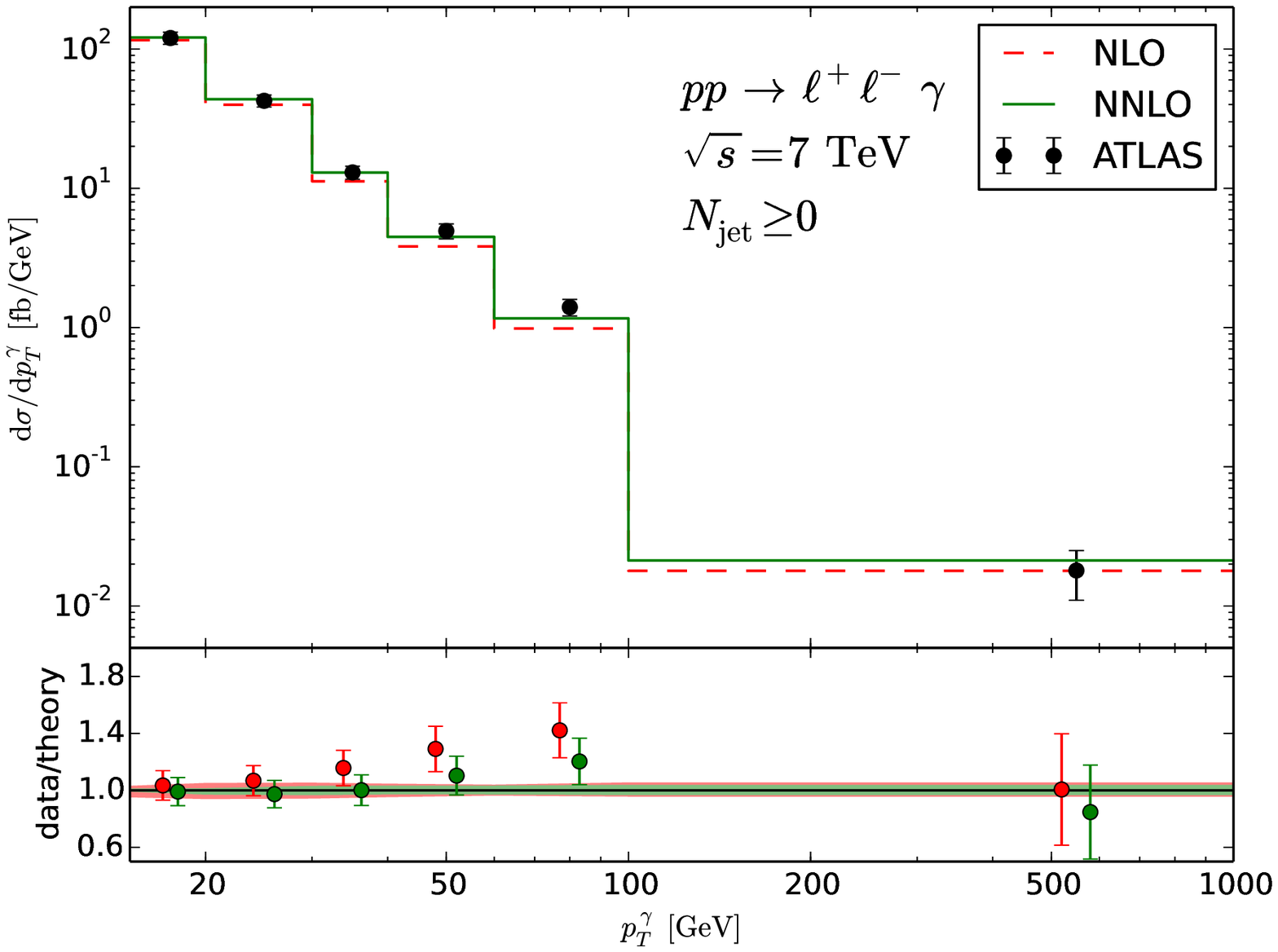}
  \hspace{-1em}
  \includegraphics[width=0.5\columnwidth]{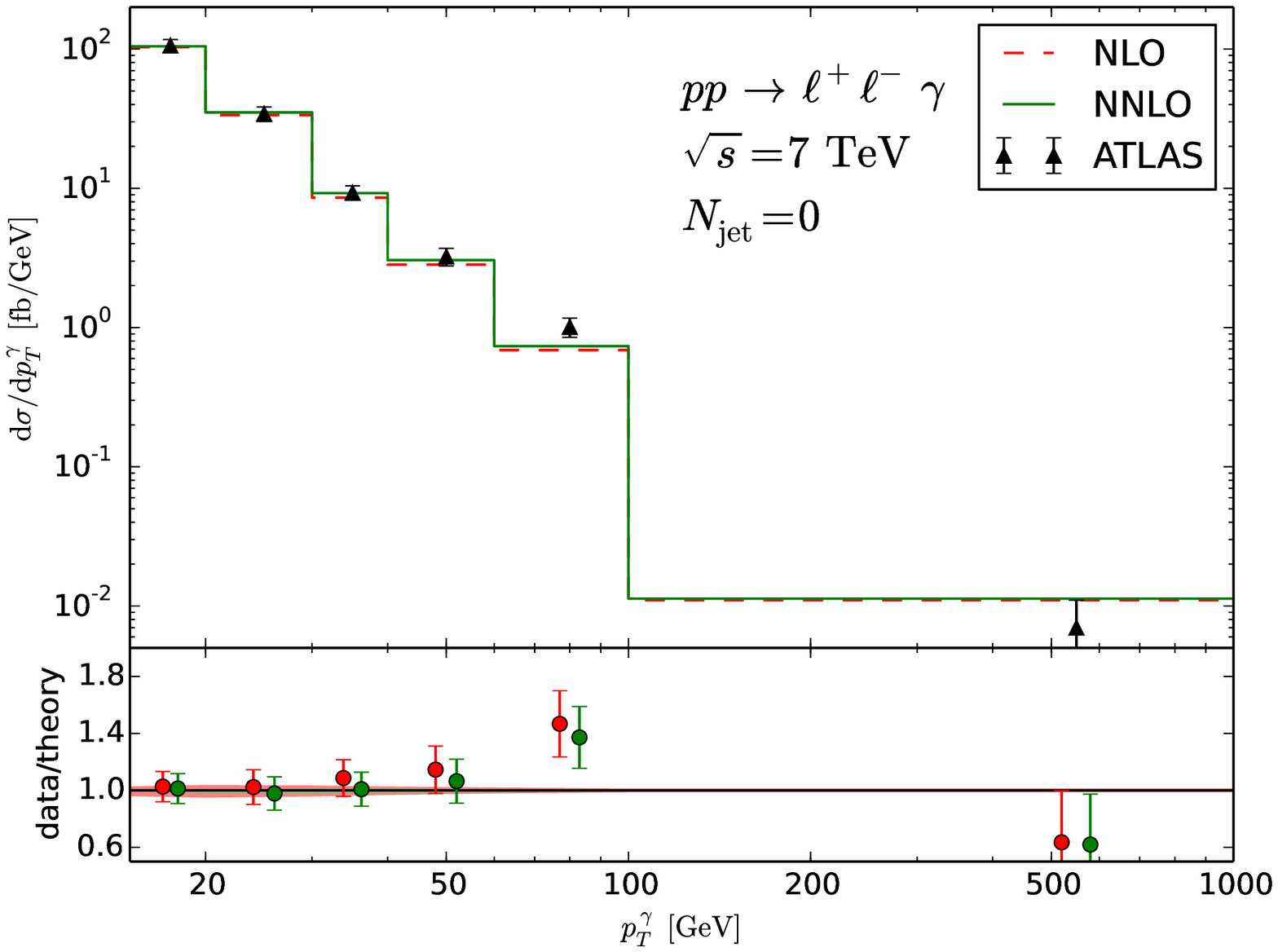}
  \caption{Photon transverse momentum distribution in the inclusive (left) and exclusive case (right) at NLO (red, dashed) 
and NNLO (green, solid) compared to ATLAS data. The lower panel shows the data/theory ratio for both theory preditions. 
In the upper panel, only experimental uncertainties are shown. The lower panel also shows theoretical uncertainty estimates from scale variations.}
  \label{fig:atlas_7_pT}
\end{figure}

Beyond the cross section in the fiducial region, ATLAS has also provided the measured cross section differential in the photon transverse momentum. 
A comparison of the resulting distribution with our theoretical NLO and NNLO predictions is displayed in \reffi{fig:atlas_7_pT} for both the inclusive 
and the exclusive case. In particular at transverse momenta $\pT^\gamma\leq 100\,\GeV$, the inclusion of NNLO corrections tends to improve the 
agreement between data and theory.
The comparison of the theoretical predictions to the data in the high transverse-momentum region $\pT^\gamma > 100\,\GeV$ is quite delicate. First, the experimental uncertainty in this region is quite large. Then,
EW corrections are expected to become sizable and negative due to large Sudakov logarithmic contributions~\cite{Hollik:2004tm,Accomando:2005ra}.

\begin{figure}[tp]
  \centering
  \includegraphics[width=0.5\columnwidth]{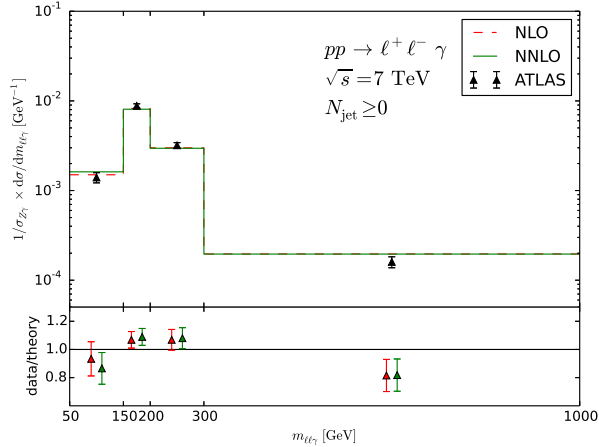}
  \caption{The invariant-mass distribution of the $\ell^+\ell^-\gamma$ system at NLO (red, dashed) and NNLO (green, solid), normalized to the fiducial cross section calculated at the repsective order, is compared to ATLAS data. The lower panel shows the data/theory ratio. Only experimental uncertainties are shown.}
  \label{fig:atlas_7_m}
\end{figure}

In \reffi{fig:atlas_7_m} we compare the NLO and NNLO predictions for the invariant-mass distribution 
of the $\ell^+\ell^-\gamma$ system with the distribution provided by ATLAS in \citere{Aad:2013izg}.
For this measurement, ATLAS increases the transverse momentum cut on the photon to $\pT^\gamma>40\,\GeV$:
the corresponding cross sections are reported in \refta{tab:zll_results_hard}. 
The relative impact of radiative corrections is 79\% and 17\%
when going from LO to NLO and from NLO to NNLO, respectively.
We conclude that the corrections are significantly larger compared
to the case in which the soft $\pT^\gamma$ cut ($\pT^\gamma>15\,\GeV$) is applied.
As the $m_{\ell^+\ell^-\gamma}$ differential cross section in \reffi{fig:atlas_7_m} is normalized by the fiducial cross section, 
sizeable NNLO corrections are visible only in the first bin, where the agreement with data is slightly improved. 
This implies that the NNLO/NLO ratio is almost constant for larger invariant masses.

\begin{figure}[tp]
  \centering
  \includegraphics[width=0.5\columnwidth]{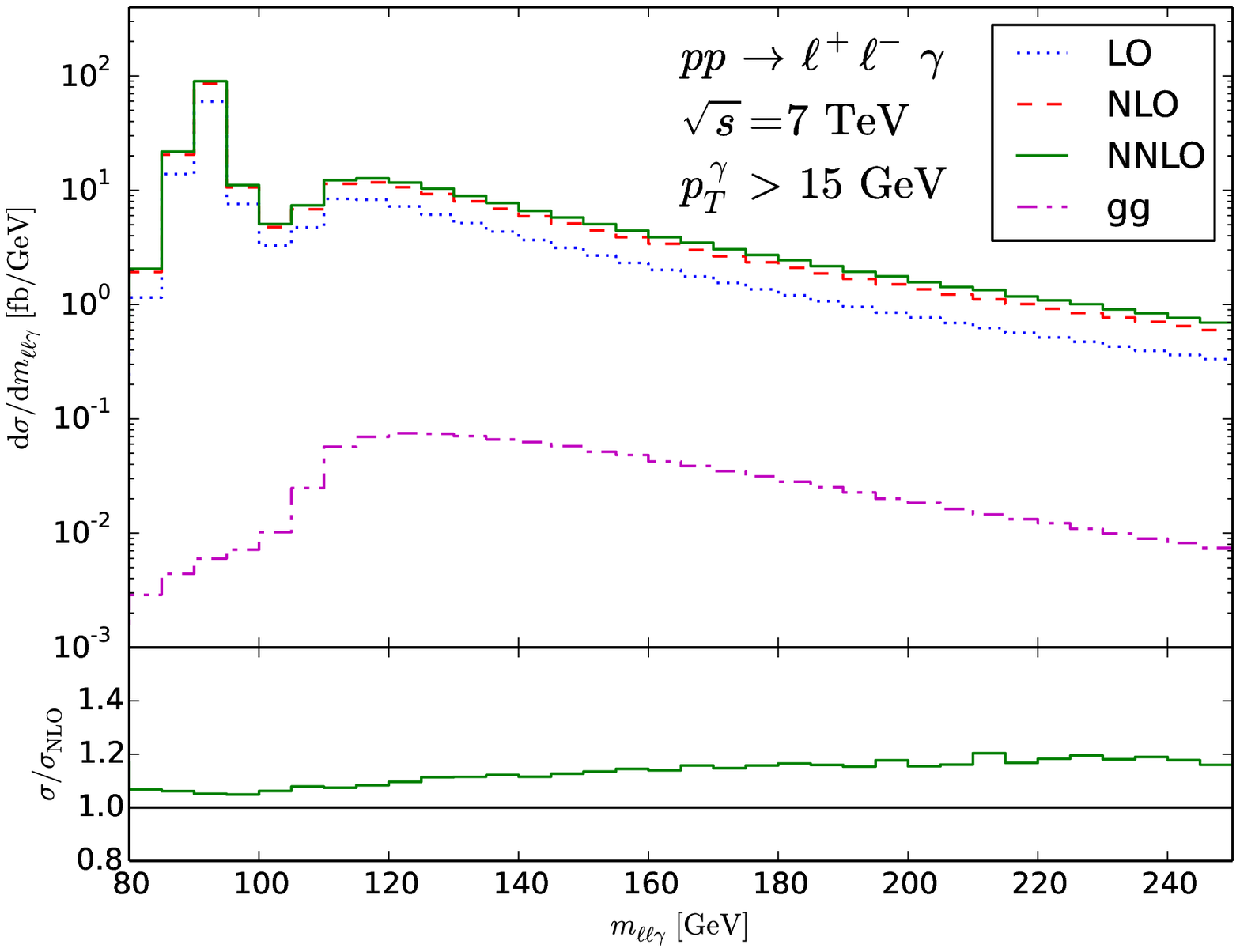}
  \hspace{-1em}
  \includegraphics[width=0.5\columnwidth]{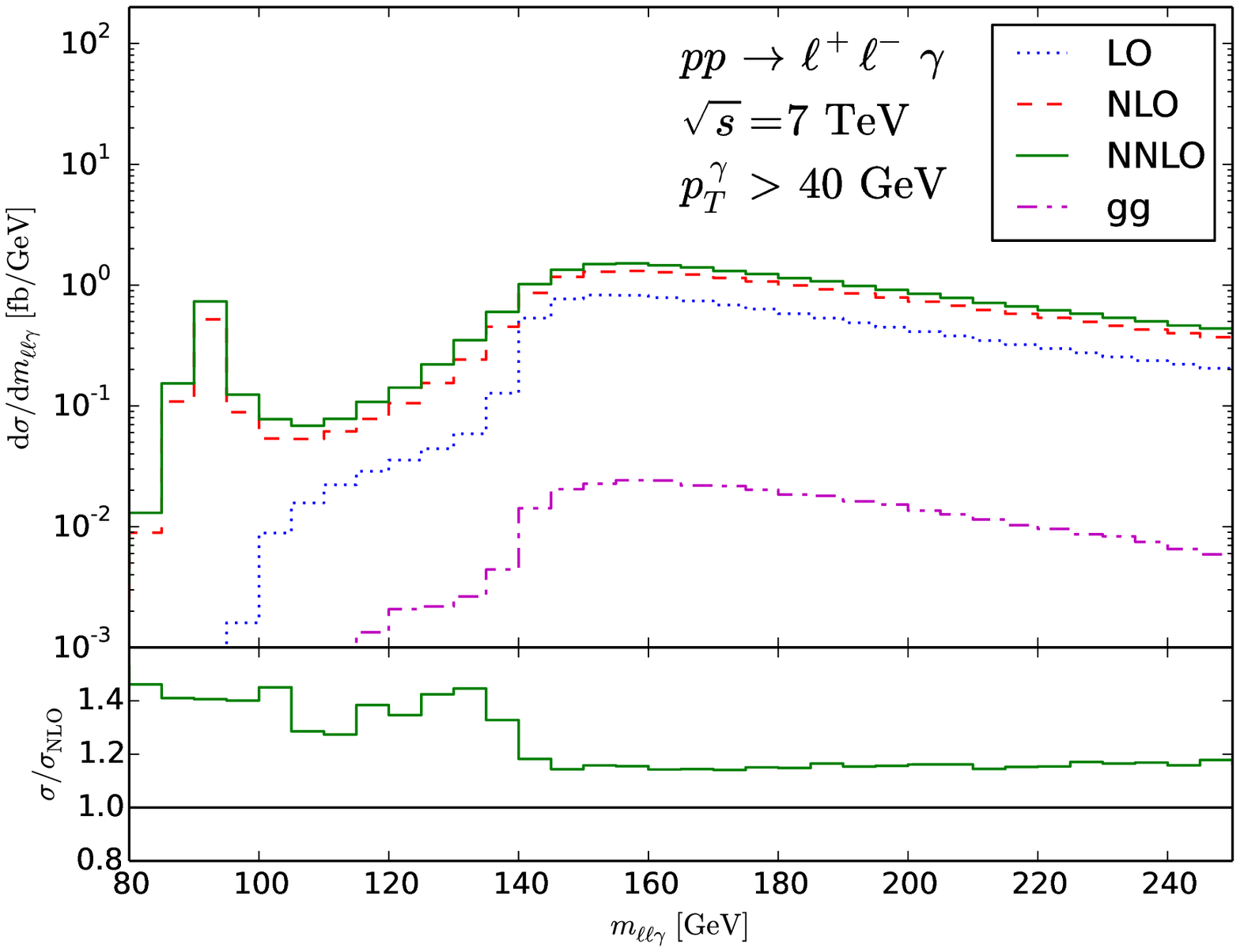}
  \caption{Invariant mass distribution of the $\ell^+\ell^-\gamma$ system at LO (blue, dotted), NLO (red, dashed) and 
NNLO (green, solid) for the setup with $\pT^\gamma>15\,\GeV$ (left) and the setup with $\pT^\gamma>40\,\GeV$ (right). 
The loop-induced gluon fusion contribution is also shown (pink, dash-dotted). The lower panel shows the NNLO/NLO ratio.}
  \label{fig:atlas_7_m_theory}
\end{figure}

The more pronounced higher-order corrections in the case in which a hard $\pT^\gamma$ cut is applied
can be understood by
studying the $\ell^+\ell^-\gamma$ invariant mass distribution in a finer binning, 
which is shown in \reffi{fig:atlas_7_m_theory} for both the soft and the hard $\pT^\gamma$ cuts. 
When the soft $\pT^\gamma$ cut is applied, the relative impact of the NNLO corrections is small in the region around the $Z\to \ell^+\ell^-\gamma$ peak, 
where the fiducial cross section receives its dominant contribution, and then slowly increases with the invariant mass. 
When the hard $\pT^\gamma$ cut is applied, the $Z\to \ell^+\ell^-\gamma$ peak is not populated at all at LO as
the applied cuts produce a lower bound at $m_{\ell^+\ell^-\gamma}\approx 97\,\GeV$ in LO kinematics.
The region below the boundary contributes sizably to the cross section, but is only
populated beyond LO, i.e.\ in this region the NLO computation provides actually only a LO prediction.
Hence the NNLO predictions effectively correspond to the first perturbative correction,
with a comparably large K factor of about $1.4$.
The lower bound on $m_{\ell^+\ell^-\gamma}$ for LO kinematics also exists with the soft $\pT^\gamma$ cut, namely at $m_{\ell^+\ell^-\gamma} \approx 66\,\GeV$. 
However, in this case the $Z\to \ell^+\ell^-\gamma$ peak is populated already at LO, and the region below the cut
does not significantly affect the fiducial cross section. 
As already expected from \reffi{fig:atlas_7_m}, the NNLO/NLO ratio in the hard $\pT^\gamma$ case
is almost independent 
of $m_{\ell^+\ell^-\gamma}$ above $m_{\ell^+\ell^-\gamma}\approx 140\,\GeV$.

\reffi{fig:atlas_7_m_theory} also shows the contribution from the loop-induced gluon fusion process, 
which, as explained in \refse{sec:outline}, respresents a finite and gauge invariant subcontribution 
to the full NNLO result. This contribution is often argued to be potentially sizable due to the large gluon luminosities at the LHC. In our calculation, however, 
the gluon fusion contribution turns out to be small. It amounts only to around $6 (9)\%$ of the full 
$\mathcal{O}\left(\as^2\right)$ 
correction and, correspondingly, to less than $1(2)\%$ of the total 
fiducial cross section in the soft and the hard $\pT^\gamma$ case, respectively.

\subsection{\texorpdfstring{$pp\to \nu_\ell\overline{\nu}_\ell\gamma$}{nunugamma production}}
\label{sec:nunugam}

In the $pp\to Z\gamma\to\nu_\ell\overline{\nu}_\ell\gamma$ analysis for proton--proton collisions 
at $\sqrt{s}=7\,\TeV$, we again use the selection criteria applied by ATLAS~\cite{Aad:2013izg}:
compared to the $pp\to \ell^+\ell^-\gamma$ analysis, the transverse-momentum cut on the photon is 
made harder ($\pT^\gamma>100\,\GeV$), and a cut on the missing transverse momentum, i.e.\ the vectorial
sum of the neutrino momenta, $\pT^{\nu\bar\nu}>90\,\GeV$, is imposed. 
The jet algorithm, the photon isolation and all other event-selection criteria
are the same as in the $pp\to \ell^+\ell^-\gamma$ analysis, if applicable. 
 
In the $\sqrt{s}=8\,\TeV$ analysis, both the photon transverse-momentum and the missing transverse-momentum cuts are 
increased to $\pT^\gamma>130\,\GeV$ and $\pT^{\nu{\bar \nu}}>100\,\GeV$, respectively.
As in the $Z\gamma\to\ell^+\ell^-\gamma$ analysis
the rapidity acceptance for 
jets is slightly increased to $|\eta^{\rm jet}|<4.5$.
The cuts are summarized in \refta{tab:znunu_cuts}.

\begin{table}[tp]
\begin{center}
\begin{tabular}{lcc}
 & \hspace*{3em}$\sqrt{s}=7\,\TeV$\hspace*{3em} & \hspace*{3em}$\sqrt{s}=8\,\TeV$\hspace*{3em}%
\Bstrut
\\
\hline

\Tstrut
Neutrinos & $\pT^{\nu\bar\nu}>90\,\GeV$ & $\pT^{\nu\bar\nu}>100\,\GeV$%
\Bstrut
\\
\hline

\Tstrut
Photon
& $\pT^\gamma>100\,\GeV$ & $\pT^\gamma>130\,\GeV$%
\Bstrut
\\
& \multicolumn{2}{c}{$|\eta^\gamma|<2.37$}%
\Bstrut
\\
& \multicolumn{2}{c}{Frixione isolation with $\varepsilon_\gamma=0.5$, $R=0.4$, $n=1$}%
\Bstrut
\\
\hline

\Tstrut
Jets
& \multicolumn{2}{c}{$\pT^{\rm jet}>30\,\GeV$}%
\Bstrut\\
& $|\eta^{\rm jet}|<4.4$ & $|\eta^{\rm jet}|<4.5$%
\Bstrut
\\
& \multicolumn{2}{c}{$N_{\rm jet}\geq 0$ (inclusive) or $N_{\rm jet}=0$ (exclusive)}%
\Bstrut
\\
\hline

\Tstrut
Separation
& \multicolumn{2}{c}{$\Delta R(\gamma,{\rm jet})>0.3$}%
\Bstrut
\\
\hline
\end{tabular}

\end{center}
\caption{Event selection criteria for $pp\to \nu_\ell\overline{\nu}_\ell\gamma$.}
\label{tab:znunu_cuts}
\end{table}

The predicted cross sections at LO, NLO and NNLO can be found in \refta{tab:znunu_results}. 
The results are presented for a single neutrino species and thus have to be multiplied by a factor of three to 
obtain the 
complete $\nu{\bar \nu}\gamma$ cross section.
In the inclusive case, i.e.\ for $N_{\mathrm{jet}}\geq 0$, we find relatively large NLO corrections of 
around 57\% and 68\% and NNLO corrections of around 12\% and 14\% at $\sqrt{s}=7\,\TeV$ and 
$\sqrt{s}=8\,\TeV$, respectively. 
The inclusive NNLO cross section prediction at $\sqrt{s}=7\,\TeV$ is in good agreement with the cross 
section measured by ATLAS. In the exclusive case, $N_{\mathrm{jet}}= 0$, the NNLO corrections are very 
small, and the scale uncertainties are reduced down
to the $1\%$ level.
We observe quite a significant discrepancy 
with respect to the ATLAS measurement for $\sqrt{s}=7\,\TeV$. The origin of this discrepancy is unclear 
to this point. As mentioned in \refse{sec:llgam} the stability of the fixed order calculation
when a jet veto is applied is challenged and the
perturbative uncertainties we find through scale variations are likely to be underestimated.
We also point out that our NLO prediction differs significantly from the NLO prediction
reported in Table VII of \citere{Aad:2013izg}, that, even using \MCFM, we are not able to reproduce.
This could be due to a large parton-to-particle correction applied on the NLO result quoted by ATLAS.
\begin{table}[tp]
\begin{center}
\begin{tabular}{c c c c c c}
$\sqrt{s}$ [TeV]
&
& $\sigma_{\mathrm{LO}}$ [fb] 
& $\sigma_{\mathrm{NLO}}$ [fb] 
& $\sigma_{\mathrm{NNLO}}$ [fb]
& $\sigma_{\mathrm{ATLAS}}$ [fb]
\Bstrut
\\ 
\hline
\Tstrut
\multirow{3}{*}{7} 
& $N_{\rm jet}\geq 0$ 
& \multirow{3}{*}{$26.27^{+0.3\%}_{-0.9\%}$}
& $41.23^{+4.1\%}_{-3.1\%}$
& $46.01^{+2.5\%}_{-2.3\%}$
& $44.3 \begin{array}{l}\\[-3ex]\scriptstyle\pm 4.3~{\rm (stat)}\\[-1ex] \scriptstyle\pm 6.7~{\rm (syst)}\\[-1ex] \scriptstyle \pm 1.7~{\rm (lumi)}\end{array}$
\Bstrut
\\
& $N_{\rm jet} = 0$ 
&  
& $29.36^{+1.2\%}_{-1.3\%}$
& $28.85^{+1.0\%}_{-0.9\%}$
& $38.7 \begin{array}{l}\\[-3ex]\scriptstyle\pm 3.3~{\rm (stat)}\\[-1ex] \scriptstyle\pm 4.3~{\rm (syst)}\\[-1ex] \scriptstyle \pm 1.3~{\rm (lumi)}\end{array}$
\Bstrut
\\
\hline
\Tstrut
\multirow{2}{*}{8}
& $N_{\rm jet}\geq 0$ 
& \multirow{2}{*}{$14.11^{+1.1\%}_{-1.5\%}$} 
& $23.66^{+4.9\%}_{-3.9\%}$
& $26.94^{+2.9\%}_{-2.7\%}$
& 
\Bstrut
\\
& $N_{\rm jet} = 0$ 
&  
& $15.09^{+1.6\%}_{-1.9\%}$
& $14.89^{+1.2\%}_{-1.0\%}$
& 
\Bstrut
\\
\hline
\end{tabular}

\end{center}
\caption{$pp\to \nu_\ell\overline{\nu}_\ell\gamma$ cross sections at LO, NLO and NNLO.
Scale uncertainties are evaluated as in \refta{tab:zll_results}.
The numerical uncertainty of the NNLO prediction from statistical error and finite $r_{\mathrm{cut}}$ is conservatively estimated to be about $0.5\%$.
The last column provides the corresponding result by ATLAS.}
\label{tab:znunu_results}
\end{table}

\begin{figure}[tp]
  \centering
  \includegraphics[width=0.5\columnwidth]{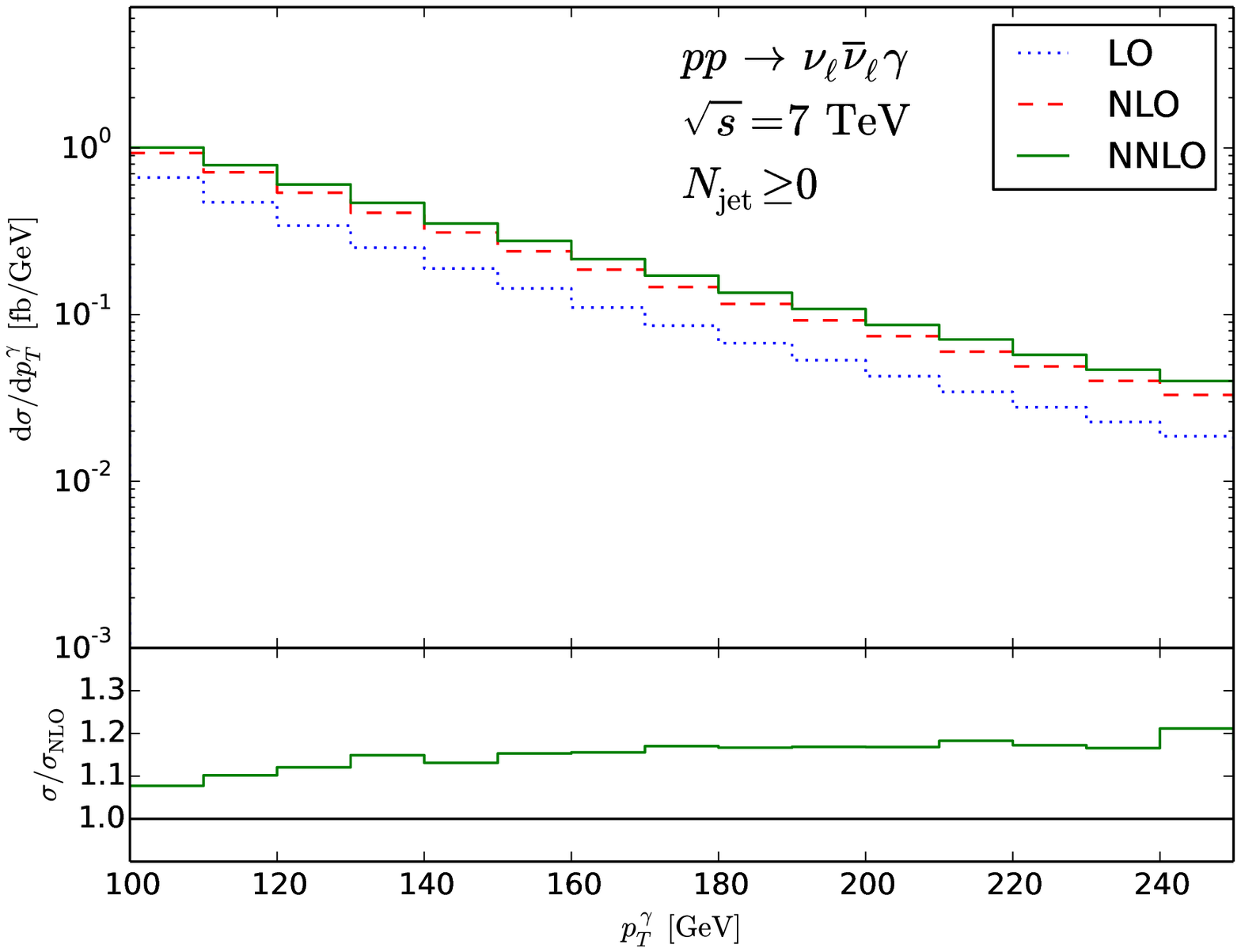}
  \hspace{-1em}
  \includegraphics[width=0.5\columnwidth]{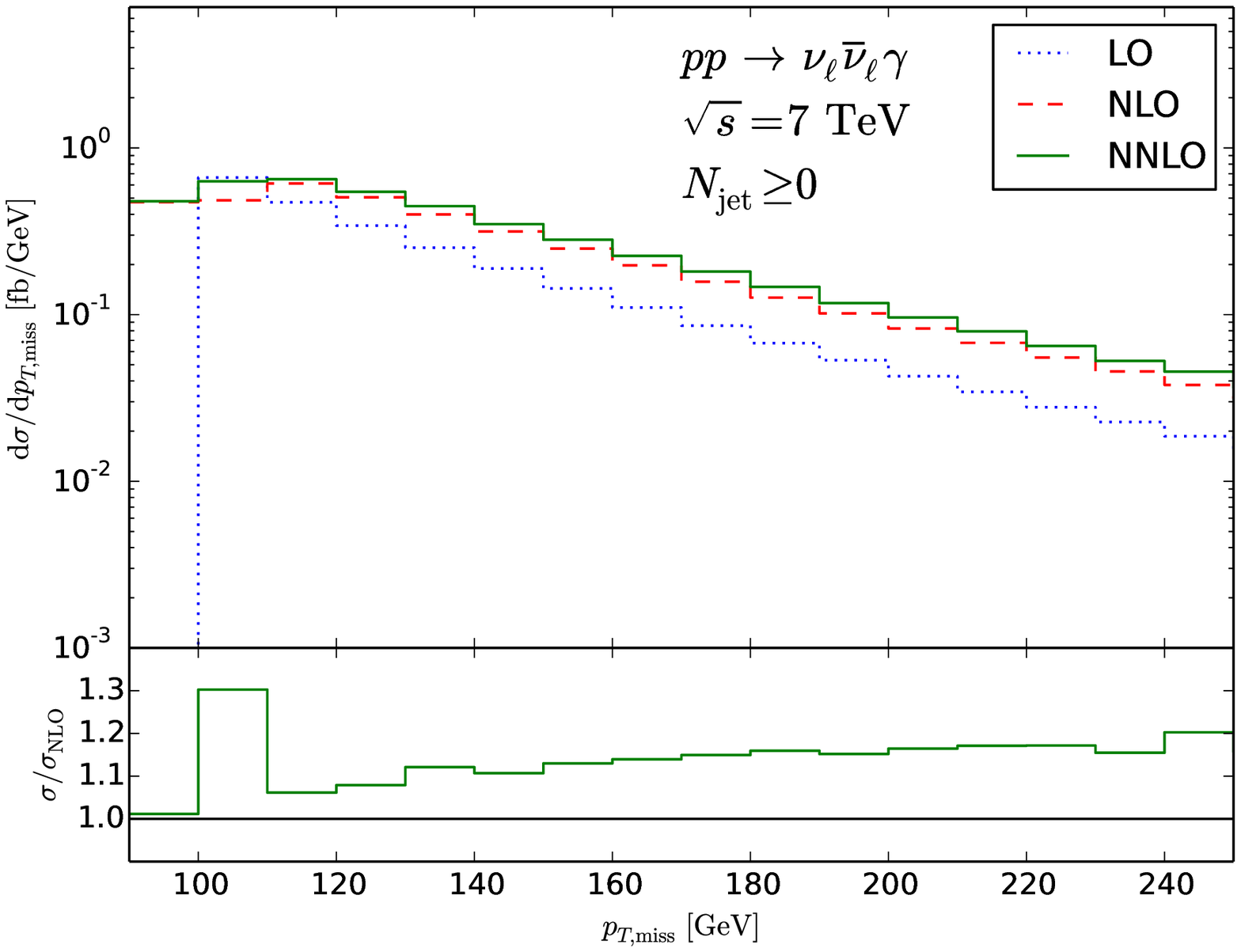}
  \caption{Photon transverse momentum (left) and missing transverse momentum (right) distribution for 
$pp\to \nu_\ell\overline{\nu}_\ell\gamma$ at LO (blue, dotted), NLO (red, dashed) and NNLO (green, solid). 
The lower panel shows the NNLO/NLO ratio.}
  \label{fig:atlas_7_znunu_theory}
\end{figure}

\reffi{fig:atlas_7_znunu_theory} shows the photon transverse-momentum and the missing 
transverse-momentum distributions. These distributions are identical for Born kinematics due to momentum conservation, 
so the difference results purely from real-radiation corrections. 
Above the photon transverse-momentum cut of $\pT^\gamma>100\,\GeV$, the difference between the two distributions is very small. 
Below a missing transverse momentum of $100\,\GeV$, the cross section is only non-vanishing starting from the NLO. \reffi{fig:atlas_7_znunu_theory} shows a perturbative instability around $p_{T,{\rm miss}}\approx 100$ GeV.
This instability originates from the incomplete cancellation of virtual and real corrections close to the phase
space boundary
(see \citere{Catani:1997xc} for a discussion of this phenomenon.) 
This class of singularities is integrable and does not alter the inclusive cross section, but would require a resummed computation to achieve a reliable differential prediction close to the boundary.

We can also study the transverse-mass distribution of the $\nu{\bar \nu}\gamma$ system, defined as
\begin{align}
 \left(\mT^{\nu\nu\gamma}\right)^2 = \left(\left|\vec{p}_T^{\,\gamma}\right|+E_T^{\mathrm{miss}}\right)^2-\left|\vec{p}_T^{\,\gamma}+\vec{E}_T^{\textrm{miss}}\right|^2.
\end{align}

\begin{figure}[tp]
  \centering
  \includegraphics[width=0.5\columnwidth]{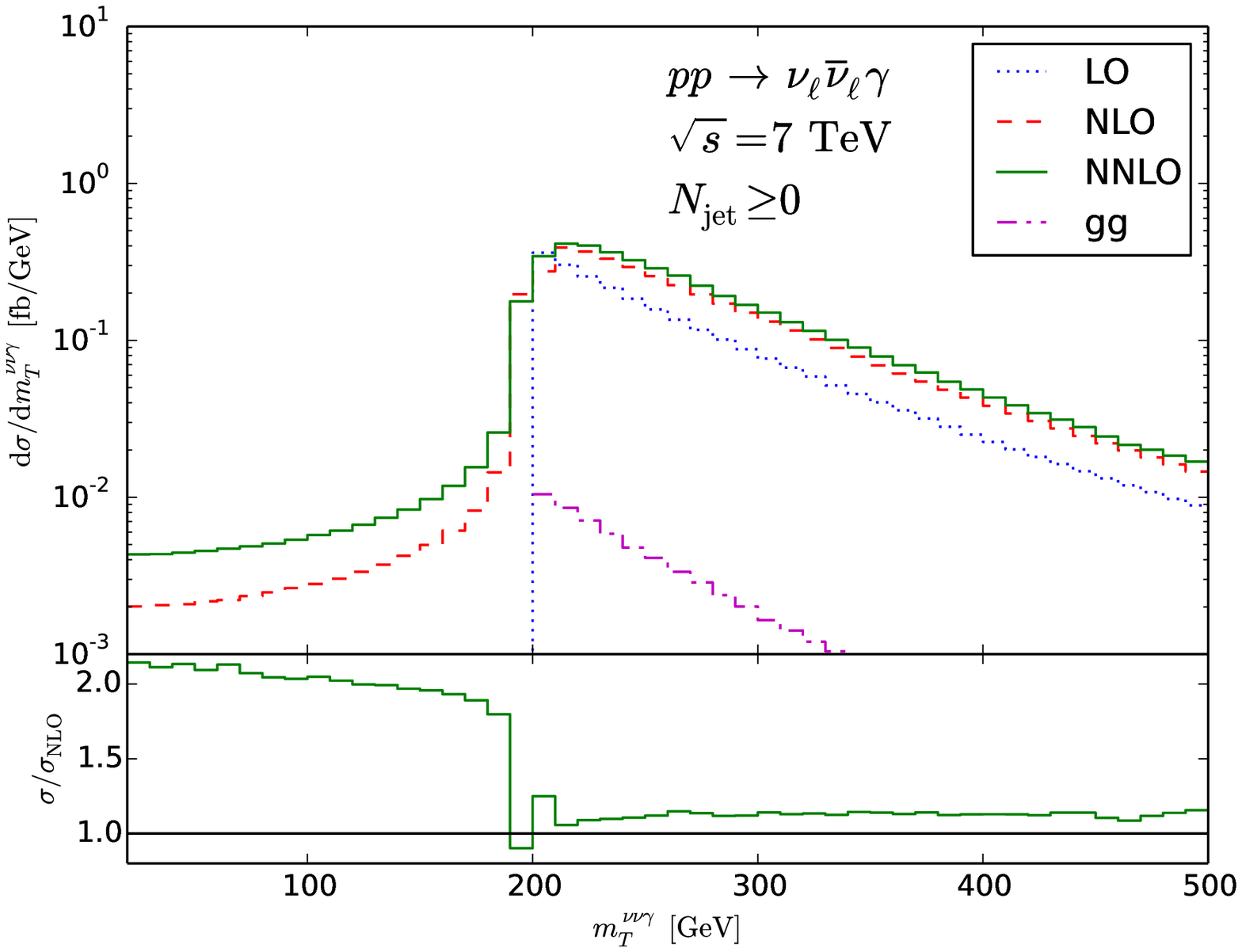}
  \hspace{-1em}
  \includegraphics[width=0.5\columnwidth]{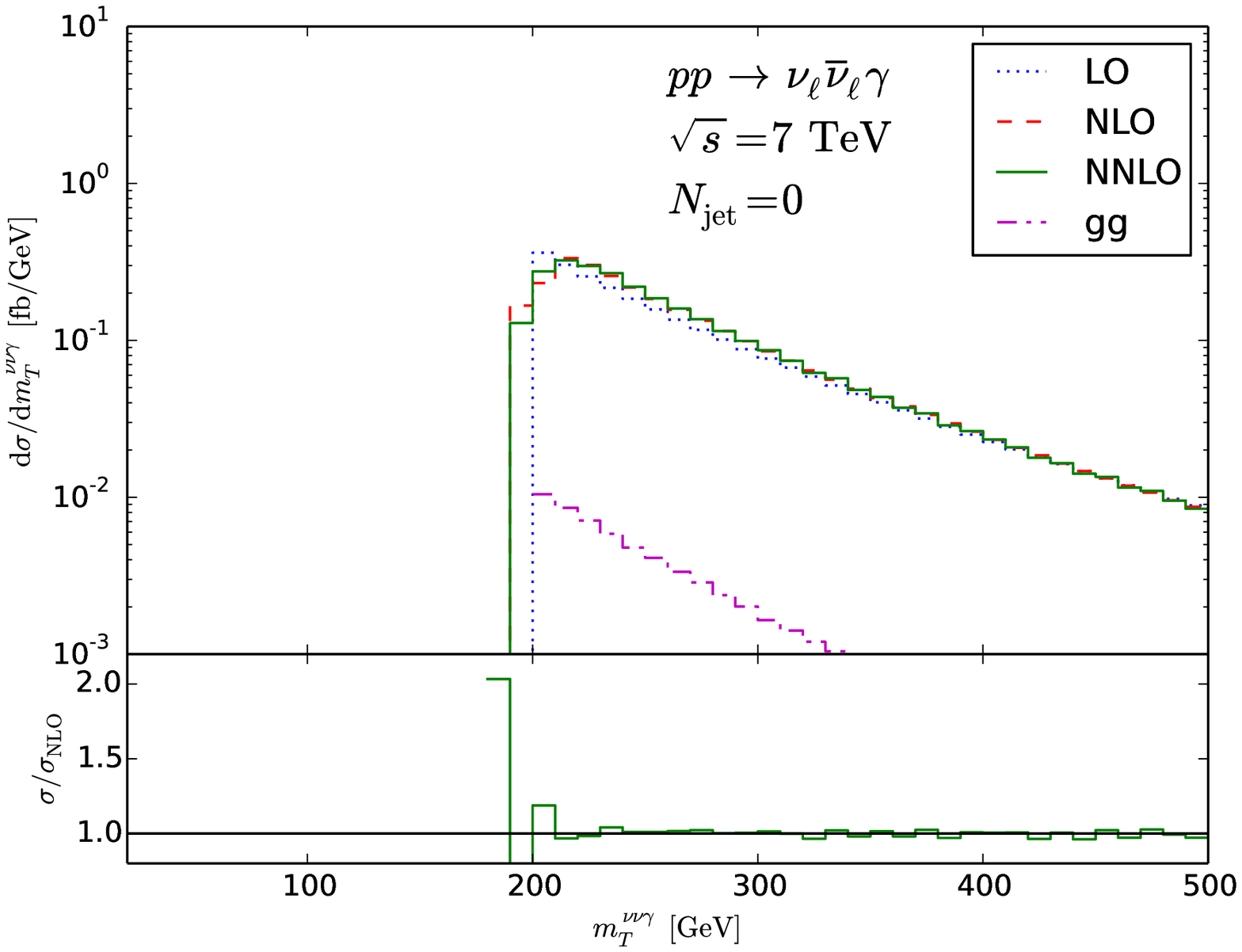}
  \caption{Transverse-mass distribution of the $\nu_\ell\overline{\nu}_\ell\gamma$ system in the inclusive (left) and exclusive case (right) at LO (blue, dotted), NLO (red, dashed) and NNLO (green, solid). The loop-induced gluon fusion contribution is also shown (pink, dash-dotted). The lower panel shows the NNLO/NLO ratio.}
  \label{fig:atlas_7_znunu_mT_theory}
\end{figure}

\reffi{fig:atlas_7_znunu_mT_theory} shows the transverse-mass distribution in the inclusive (left) and exclusive case (right). Transverse masses below $\mT^{\nu\nu\gamma}\approx 200\,\GeV$ are not allowed in LO kinematics and thus are only populated by real corrections starting from the NLO. This leads to an increased impact of the NNLO corrections in the region $\mT^{\nu\nu\gamma}<200\,\GeV$ in the inclusive case, with corrections of about 100\% compared to the NLO prediction. When applying a jet veto, this effect vanishes almost completey, indicating that relatively hard QCD radiation is necessary to overcome the LO kinematics phase space constraint. In fact, with a $30\,\GeV$ jet veto present, the real radiation does only populate the phase space down to $\mT^{\nu\nu\gamma}\approx 187\,\GeV$ at NLO and down to  $\mT^{\nu\nu\gamma}\approx 175\,\GeV$ at NNLO.

\reffi{fig:atlas_7_znunu_mT_theory} also shows the contribution from gluon fusion separately, which again is quite small and amounts to less than 2\% of the fiducial cross section in the inclusive case and about 3\% in the exclusive case.

\subsection{\texorpdfstring{$pp\to \ell\nu_\ell \gamma$}{Wgamma production}}
\label{sec:wgam}

We now present results for $pp\to \ell\nu_\ell \gamma$
at $\sqrt{s}=7\,\TeV$ and $8$ TeV.
We again use the event selection criteria adopted in the ATLAS analysis~\cite{Aad:2013izg}.
This set of cuts is identical to that used in the $pp\to \ell^+\ell^-\gamma$ analysis, 
apart from the fact that the cut on the invariant mass of the leptons is replaced with
a cut on the missing transverse momentum (which coincides with the 
transverse momentum of the neutrino from the $W$ decay) of $p_{T}^{\nu}>35\,\GeV$.
As in the case of $Z\gamma$
in our $\sqrt{s}=8\,\TeV$ analysis the rapidity of the jets is required to be $|\eta^{\rm jet}|<4.5$. 
A summary of all cuts and event selection criteria can be found in \refta{tab:wgamma_cuts}.

\begin{table}[tp]
\begin{center}
\begin{tabular}{lcc}
& \hspace*{3em}$\sqrt{s}=7\,\TeV$\hspace*{3em} & \hspace*{3em}$\sqrt{s}=8\,\TeV$\hspace*{3em}%
\Bstrut
\\
\hline

\Tstrut
Lepton
& \multicolumn{2}{c}{$\pT^\ell>25\,\GeV$}
\Bstrut
\\
& \multicolumn{2}{c}{$|\eta|<2.47$}
\Bstrut
\\
\hline
\Tstrut

Neutrino
& \multicolumn{2}{c}{$\pT^\nu>35\,\GeV$}
\Bstrut
\\
\hline

\Tstrut
Photon
& \multicolumn{2}{c}{$\pT^\gamma>15\,\GeV$ (soft $\pT^\gamma$ cut) or $\pT^\gamma>40\,\GeV$ (hard $\pT^\gamma$ cut)}
\Bstrut
\\
& \multicolumn{2}{c}{$|\eta^\gamma|<2.37$}
\Bstrut
\\
& \multicolumn{2}{c}{Frixione isolation with $\varepsilon_\gamma=0.5$, $R=0.4$, $n=1$}
\Bstrut
\\
\hline

\Tstrut
Jets 
& \multicolumn{2}{c}{anti-$k_\mathrm{T}$ algorithm with $D=0.4$}
\Bstrut
\\
& \multicolumn{2}{c}{$\pT^{\rm jet}>30\,\GeV$}
\Bstrut
\\
& $|\eta^{\rm jet}|<4.4$ & $|\eta^{\rm jet}|<4.5$
\Bstrut
\\
& \multicolumn{2}{c}{$N_{\rm jet}\geq 0$ (inclusive) or $N_{\rm jet}=0$ (exclusive)}
\Bstrut
\\
\hline

\Tstrut
Separation
& \multicolumn{2}{c}{$\Delta R(\ell,\gamma)>0.7$}
\Bstrut
\\
& \multicolumn{2}{c}{$\Delta R(\ell/\gamma,{\rm jet})>0.3$}
\Bstrut
\\
\hline
\end{tabular}

\end{center}
\caption{$W^\pm\left(\to \nu_\ell\ell\right)\gamma$ cuts and event-selection criteria.}
\label{tab:wgamma_cuts}
\end{table}

All results in the following will be presented summed over the $W$ charges, i.e.\ we combine the processes 
$pp\to W^+\gamma$ and $pp\to W^-\gamma$, to facilitate the comparison with experimental data. 
The predicted fiducial cross sections both for the inclusive and the exclusive case can be found in 
\refta{tab:wgam_results}.
In the inclusive case, the NLO corrections are quite large, and amount to about $136$--$143\%$. The NNLO 
corrections increase the NLO result by $19$--$20\%$. The impact of higher order corrections is thus much larger than in the case of 
$Z\gamma$ production. We will come back to this point in \refse{sec:disc}.

\refta{tab:wgam_results} also shows the cross sections measured by ATLAS. The measurement of the inclusive cross sections
shows a $2\sigma$ excess with respect to the NLO prediction, which is reduced to well below $1\sigma$ when including the NNLO corrections.

The impact of QCD corrections at NLO and NNLO is reduced to
$60\%$ and $7\%$, respectively, when the jet veto is applied ($N_{\rm jet}=0$).
As discussed in \refse{sec:llgam}, such an effect is expected and apparently leads to a more stable perturbative prediction,
but also to the possible need of more conservative procedures to estimate perturbative uncertainties.
In the exclusive case, the excess of the measured fiducial cross sections over the theoretical prediction
is reduced from $1.6\sigma$ 
to $1.2\sigma$ when going from NLO to NNLO.
We note that the scale variations at NLO significantly underestimate the impact of the NNLO corrections, 
in particular in the inclusive case. 

\begin{table}[tp]
\begin{center}
\begin{tabular}{c c c c c c}
$\sqrt{s}$ [TeV]
& 
& $\sigma_{\textrm{LO}}$ [pb] 
& $\sigma_{\textrm{NLO}}$ [pb] 
& $\sigma_{\textrm{NNLO}}$ [pb]
& $\sigma_{\textrm{ATLAS}}$ [pb]
\Bstrut
\\ 
\hline
\Tstrut
\multirow{3}{*}{7} 
& $N_{\rm jet}\geq 0$ 
& \multirow{3}{*}{$0.8726^{+6.8\%}_{-8.1\%}$} 
& $2.058^{+6.8\%}_{-6.8\%}$ 
& $2.453^{+4.1\%}_{-4.1\%}$
& $2.77 \begin{array}{l}\\[-3ex]\scriptstyle\pm 0.03~{\rm (stat)}\\[-1ex] \scriptstyle\pm 0.33~{\rm (syst)}\\[-1ex] \scriptstyle \pm 0.14~{\rm (lumi)}\end{array}$
\Bstrut
\\
& $N_{\rm jet} = 0$ 
&  
& $1.395^{+5.2\%}_{-5.8\%}$
& $1.493^{+1.7\%}_{-2.7\%}$
& $1.76 \begin{array}{l}\\[-3ex]\scriptstyle\pm 0.03~{\rm (stat)}\\[-1ex] \scriptstyle\pm 0.21~{\rm (syst)}\\[-1ex] \scriptstyle \pm 0.08~{\rm (lumi)}\end{array}$
\Bstrut
\\
\hline
\Tstrut
\multirow{2}{*}{8}
& $N_{\rm jet}\geq 0$ 
& \multirow{2}{*}{$0.9893^{+7.7\%}_{-9.1\%}$} 
& $2.401^{+7.4\%}_{-7.4\%}$ 
& $2.884^{+4.1\%}_{-4.3\%}$
& 
\Bstrut
\\
& $N_{\rm jet} = 0$ 
&  
& $1.587^{+5.5\%}_{-6.3\%}$
& $1.691^{+1.8\%}_{-2.9\%}$
& 
\Bstrut
\\
\hline
\end{tabular}

\end{center}
\caption{$W^\pm\left(\to \nu_\ell\ell\right)\gamma$ cross sections with the soft $\pT^\gamma$ cut ($\pT^\gamma>15\,\GeV$). 
Scale uncertainties are computed as in \refta{tab:zll_results}. 
The numerical uncertainty of the NNLO prediction from statistical error and finite $r_{\mathrm{cut}}$ is conservatively estimated to be about $0.8\%$.
The last column provides the measured cross sections provided by ATLAS.}
\label{tab:wgam_results}
\end{table}

\begin{table}[tp]
\begin{center}
\begin{tabular}{c c c c c c}
$\sqrt{s}$ [TeV]
&
& $\sigma_{\mathrm{LO}}$ [fb] 
& $\sigma_{\mathrm{NLO}}$ [fb] 
& $\sigma_{\mathrm{NNLO}}$ [fb]
\Bstrut
\\ 
\hline
\Tstrut
7
& $N_{\rm jet}\geq 0$ 
& $115.8^{+2.6\%}_{-3.7\%}$ 
& $395.9^{+9.0\%}_{-7.3\%}$ 
& $497.1^{+5.3\%}_{-4.7\%}$
\Bstrut
\\
\hline
\Tstrut
8
& $N_{\rm jet}\geq 0$ 
& $133.0^{+3.5\%}_{-4.6\%}$ 
& $478.6^{+8.4\%}_{-7.0\%}$ 
& $604.3^{+5.2\%}_{-4.5\%}$
\Bstrut
\\
\hline
\end{tabular}

\end{center}
\caption{$W^\pm\left(\to \nu_\ell\ell\right)\gamma$ cross sections with the hard $\pT^\gamma$ cut ($\pT^\gamma>40\,\GeV$). 
Scale uncertainties are computed as in \refta{tab:zll_results}.
The numerical uncertainty of the NNLO prediction from statistical error and finite $r_{\mathrm{cut}}$ is conservatively estimated to be about $0.5\%$.}
\label{tab:wgam_results_hard}
\end{table}

\begin{figure}[tp]
  \centering
  \includegraphics[width=0.5\columnwidth]{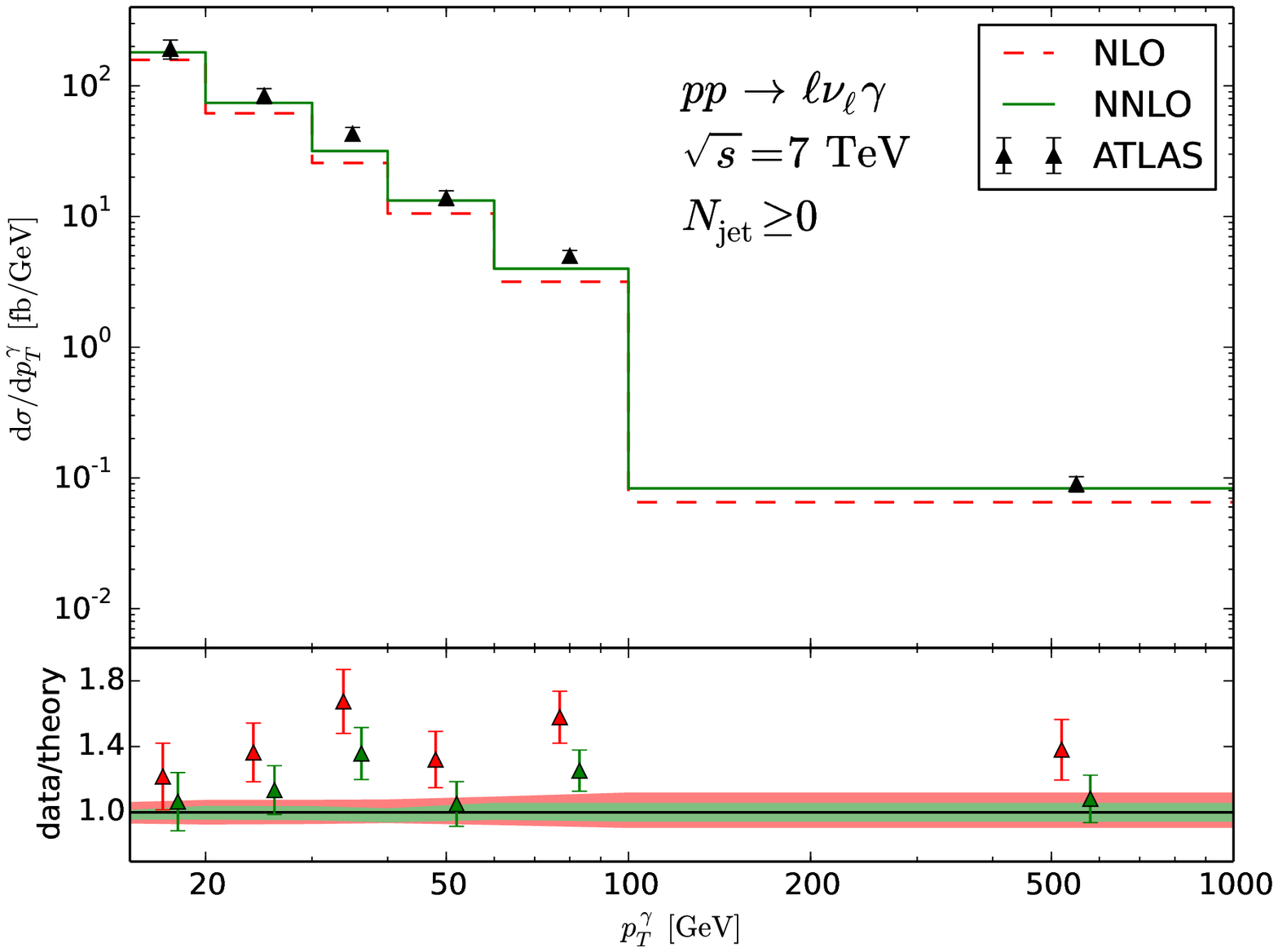}
  \hspace{-1em}
  \includegraphics[width=0.5\columnwidth]{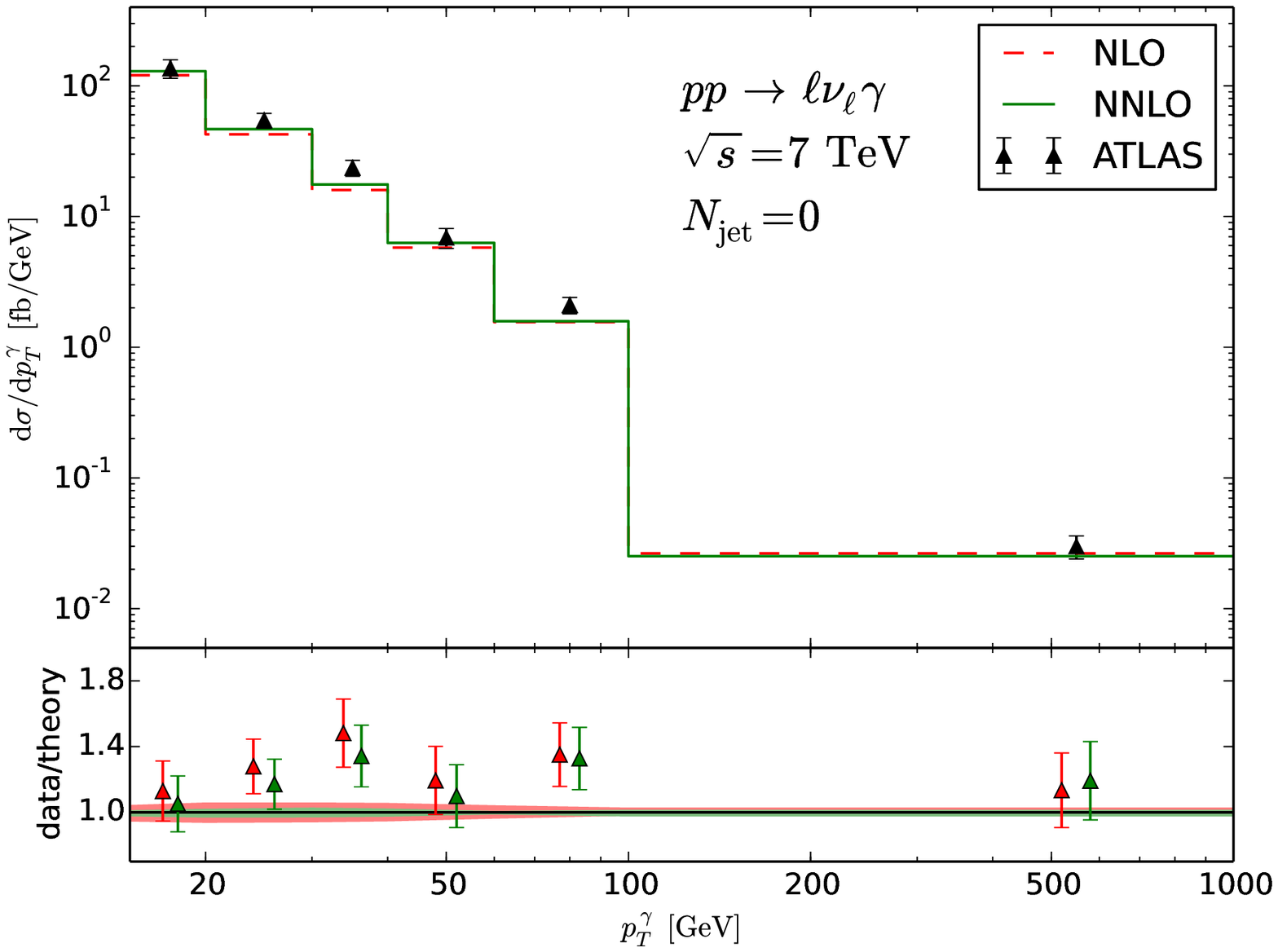}
  \caption{Photon transverse momentum distribution in the inclusive (left) and exclusive case (right) 
at NLO (red, dashed) and NNLO (green, solid) compared to ATLAS data. The lower panel shows the data/theory ratio. 
In the upper panel, only experimental uncertainties are shown. The lower panel also shows theoretical uncertainty estimates from scale variations.}
  \label{fig:wgam_atlas_7_pT}
\end{figure}

\reffi{fig:wgam_atlas_7_pT} shows the 
photon transverse-momentum distribution in comparison with 
the ATLAS measurement, both in the inclusive and in the exclusive case.
Although the experimental uncertainties are large, the agreement between data and theory 
is clearly improved when including the NNLO corrections, in particular if no veto
on jets is applied.

\begin{figure}[tp]
  \centering
  \includegraphics[width=0.5\columnwidth]{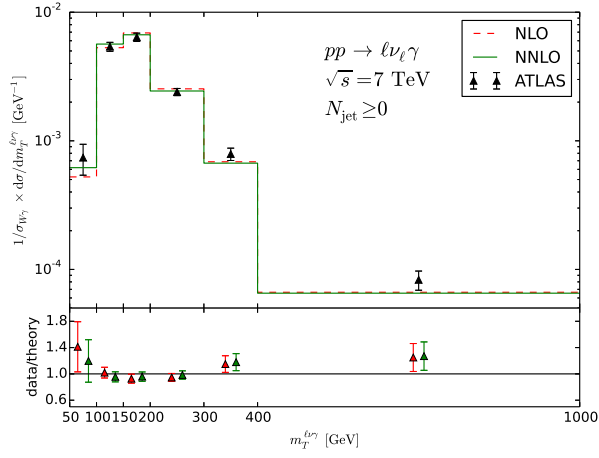}
  \caption{Transverse-mass distribution of the $\ell^\pm\nu_\ell\gamma$ system, normalized to the respective fiducial cross section 
at NLO (red, dashed) and NNLO (green, solid), compared to ATLAS data. The lower panel shows the data/theory ratio. Only experimental uncertainties are shown.}
  \label{fig:wgam_atlas_7_mT}
\end{figure}

\reffi{fig:wgam_atlas_7_mT} shows the $W\gamma$ cross section differential in the transverse mass of 
the $\ell\nu_\ell\gamma$ system, normalized by the total fiducial cross section at the respective order.
The transverse mass is defined here as
\begin{align}
\left(\mT^{\ell\nu\gamma}\right)^2 = \left(\sqrt{m_{\ell\gamma}^2+\left|\vec{p}_T^{\,\gamma}+\vec{p}_T^{\,\ell}\right|^2}+E_T^{\mathrm{miss}}\right)^2-\left|\vec{p}_T^{\,\gamma}+\vec{p}_T^{\,\ell}+\vec{E}_T^{\textrm{miss}}\right|^2.
\end{align}
The calculation is done with the hard photon transverse-momentum cut $\pT^\gamma>40\,\GeV$.
The corresponding fiducial cross sections at LO, NLO, and NNLO are reported in
\refta{tab:wgam_results_hard}.
The impact of QCD radiative corrections is $242$--$260\%$ and $26\%$ at NLO and NNLO, respectively.
In \reffi{fig:wgam_atlas_7_mT}, due to the normalization, the large overall corrections mostly cancel out, in particular at high transverse mass, 
and we observe only a slightly improved agreement with data when going from NLO to NNLO.

\begin{figure}[tp]
  \centering
  \includegraphics[width=0.5\columnwidth]{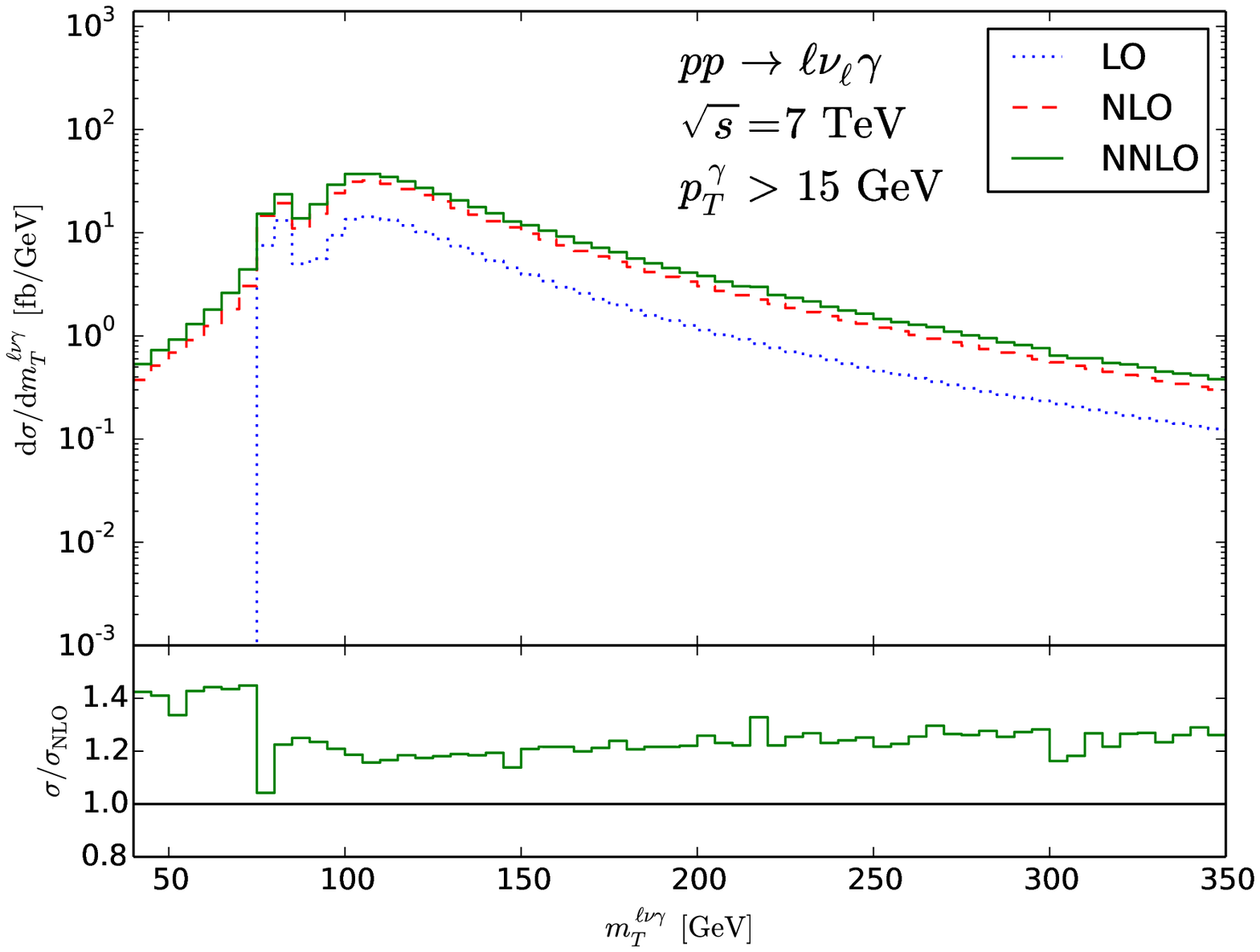}
  \hspace{-1em}
  \includegraphics[width=0.5\columnwidth]{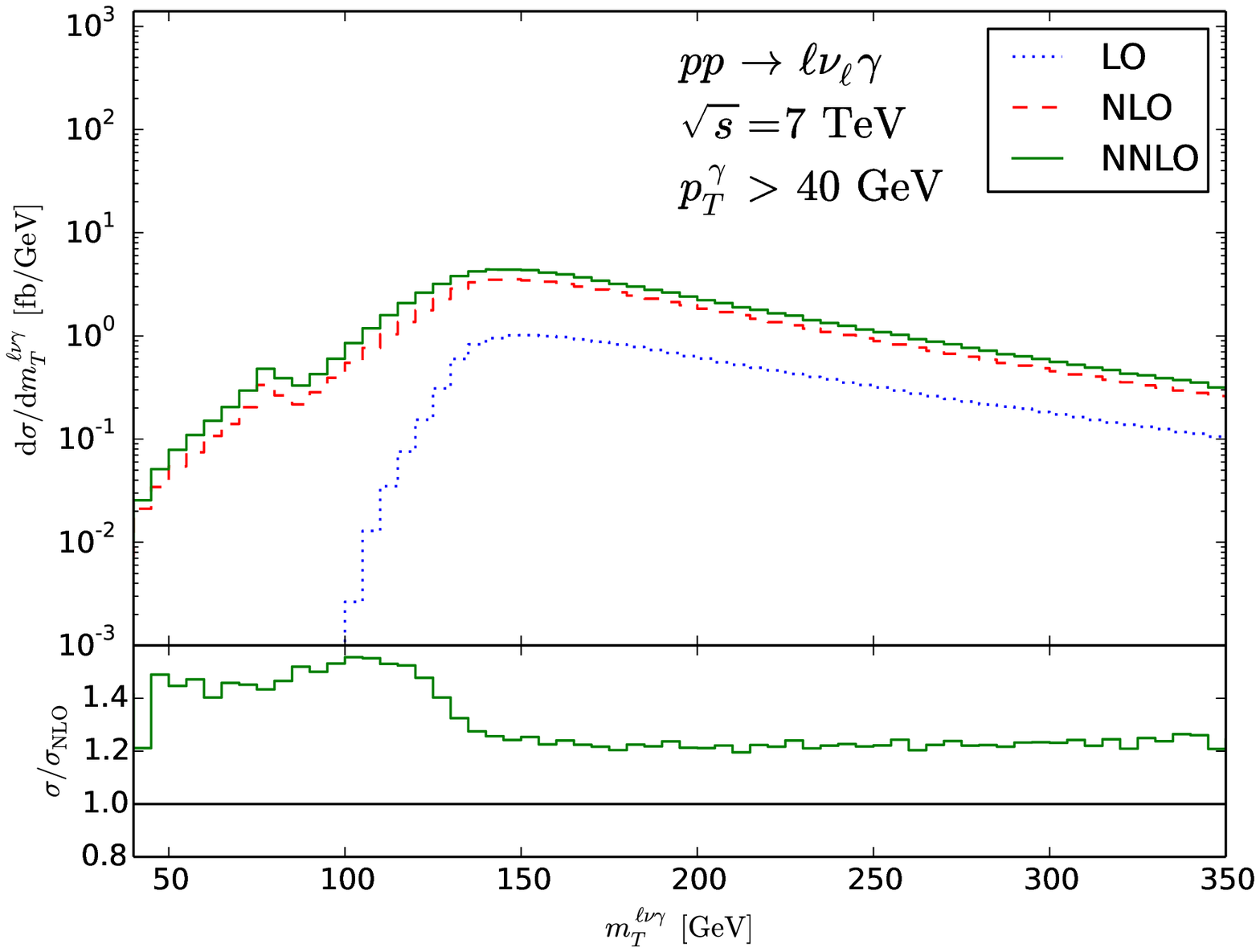}
  \caption{Transverse-mass distribution of the $\ell\nu_\ell\gamma$ system at LO (blue, dotted), 
NLO (red, dashed) and NNLO (green, solid) for $\pT^\gamma>15\,\GeV$ (left) and 
$\pT^\gamma>40\,\GeV$ (right), in the inclusive case ($N_{\rm jet}\geq 0$). The lower panel shows the NNLO/NLO ratio.}
  \label{fig:wgam_7_mT_theory}
\end{figure}

The increased relative impact of NLO and NNLO corrections when a harder $\pT^\gamma$ cut  ($\pT^\gamma>40\,\GeV$)
is applied can, in analogy to the $Z\gamma$ case (see \refse{sec:llgam}), be better understood 
by studying the transverse-mass distributions with the soft and hard $\pT^\gamma$ cut in more detail. 
The corresponding plots with a finer binning are shown in \reffi{fig:wgam_7_mT_theory}. 
When $\pT^\gamma>15$ GeV, for Born kinematics the transverse mass has a lower bound at about $\mT^{\ell\nu\gamma}\gtrsim 75\, \GeV$, 
i.e.\ below the $W\to\ell\nu_\ell\gamma$ peak. When the photon transverse-momentum cut
is increased to $40$ GeV,
the lower bound increases to $\mT^{\ell\nu\gamma}\gtrsim 100\,\GeV$,
and the $W\to\ell\nu_\ell\gamma$ peak is only populated by 
real emissions starting from the NLO, leading to large corrections in the region where the cross section is sizeable, and thus explaining the effect on the fiducial cross section.

\begin{table}[tp]
\begin{center}
\begin{tabular}{c c c c c c c}
process
& $p^\gamma_{\rm{T,cut}}$
& $\sqrt{s}$ [TeV]
& $N_{\rm{jet}}$
& $\sigma_{\mathrm{NLO}}/\sigma_{\mathrm{LO}}$
& $\sigma_{\mathrm{NNLO}}/\sigma_{\mathrm{NLO}}$
\Bstrut
\\ 
\hline

\Tstrut
$Z\left(\to \ell^+\ell^-\right)\gamma$
& \multirow{4}{*}{soft}
& \multirow{2}{*}{7}
& $N_{\rm{jet}}\geq0$
& \phantom{0}+50\%%
& \phantom{00}+8\%%
\\

& 
& 
& $N_{\rm{jet}}=0$
& \phantom{0}+27\%%
& \phantom{00}+3\%%
\Bstrut
\\

& 
& \multirow{2}{*}{8}
& $N_{\rm{jet}}\geq0$
& \phantom{0}+50\%%
& \phantom{00}+8\%%
\\

& 
& 
& $N_{\rm{jet}}=0$
& \phantom{0}+25\%%
& \phantom{00}+3\%%
\Bstrut
\\

\cline{2-6}

\Tstrut
& \multirow{2}{*}{hard}
& 7
& $N_{\rm{jet}}\geq0$
& \phantom{0}+79\%%
& \phantom{0}+17\%%
\Bstrut
\\

& 
& 8
& $N_{\rm{jet}}\geq0$
& \phantom{0}+82\%%
& \phantom{0}+18\%%
\Bstrut
\\
\hline

\Tstrut
$Z\left(\to \nu_l\overline{\nu}_l\right)\gamma$
& 
& \multirow{2}{*}{7}
& $N_{\rm{jet}}\geq0$
& \phantom{0}+57\%%
& \phantom{0}+12\%%
\\

& 
& 
& $N_{\rm{jet}}=0$
& \phantom{0}+12\%%
& \phantom{00}$-2$\%%
\Bstrut
\\

& 
& \multirow{2}{*}{8}
& $N_{\rm{jet}}\geq0$
& \phantom{0}+68\%%
& \phantom{0}+14\%%
\\

& 
& 
& $N_{\rm{jet}}=0$
& \phantom{00}+7\%%
& \phantom{00}$-1$\%%
\Bstrut
\\
\hline

\Tstrut
$W\left(\to \ell\nu_\ell\right)\gamma$
& \multirow{4}{*}{soft}
& \multirow{2}{*}{7}
& $N_{\rm{jet}}\geq0$
& +136\%%
& \phantom{0}+19\%%
\\

& 
& 
& $N_{\rm{jet}}=0$
& \phantom{0}+60\%%
& \phantom{00}+7\%%
\Bstrut
\\

& 
& \multirow{2}{*}{8}
& $N_{\rm{jet}}\geq0$
& +143\%%
& \phantom{0}+20\%%
\\

& 
& 
& $N_{\rm{jet}}=0$
& \phantom{0}+60\%%
& \phantom{00}+7\%%
\Bstrut
\\

\cline{2-6}

\Tstrut
& \multirow{2}{*}{hard}
& 7
& $N_{\rm{jet}}\geq0$
& +242\%%
& \phantom{0}+26\%%
\Bstrut
\\

& 
& 8
& $N_{\rm{jet}}\geq0$
& +260\%%
& \phantom{0}+26\%%
\Bstrut
\\
\hline
\end{tabular}

\end{center}
\caption{Summary of the relative NLO and NNLO corrections in the channels under investigation, 
$Z\left(\to \ell^+\ell^-\right)\gamma$, $Z\left(\to \nu_\ell\overline{\nu}_\ell\right)\gamma$,
 and $W^\pm\left(\to \ell^\pm\bar\nu_\ell\right)\gamma$. 
Numbers are reported for both soft and hard $\pT^\gamma$ cuts.}
\label{tab:summary_relative_corrections}
\end{table}

\subsection{The difference between $W\gamma$ and $Z\gamma$}
\label{sec:disc}

It is interesting to compare
the relative size of the NLO and NNLO corrections to the $Z\gamma$ and $W\gamma$ processes we have considered.
The results are summarized in \refta{tab:summary_relative_corrections}. It is clear that the $W\gamma$ process features much larger radiative effects with respect to the $Z\gamma$ processes.
This should be contrasted to what happens in the case of inclusive $W$ and $Z$ boson production,
where QCD radiative corrections are essentially identical~\cite{Hamberg:1990np}. It is thus the emission of the additional
photon that breaks the similarity between the charged current and the neutral current processes.

Restricting the analysis to NLO for the moment, the main source for the difference between $Z\gamma$ and $W\gamma$
can be traced back to 
the $gq$ and $g\overline{q}$ channels, which contribute a moderate, negative amount to the cross section 
in $Z\gamma$ production, but are large and positive for $W^\pm\gamma$. 
It turns out that this effect is driven by resonant $W\gamma$ contributions to the cross section, 
i.e.\ by $pp\to W(\to\ell\nu_\ell)\gamma$ topologies,
and not by $pp\to W\to\ell(\to\ell\gamma)\nu_\ell$ topologies, where the photon is emitted from the final-state lepton.
These two contributions can only be separated in a gauge-invariant way if the $W$ bosons are treated
as on-shell particles, i.e.\ in a narrow-width approximation.
By studying the LO contributions to the $Z\gamma$ and $W\gamma$ cross sections (see \reffi{fig:zgam_diagrams} and \reffi{fig:wgam_diagrams}) it turns out that in $W\gamma$ there is an
additional Feynman diagram in which the photon is radiated off the $W$ boson (see \reffi{fig:wgam_III}). This additional diagram is responsible
for a \textit{radiation zero}~\cite{Mikaelian:1979nr}, an exact zero present
in the on-shell partonic $W\gamma$ tree-level amplitude at $\cos\theta^*=1/3$, where $\theta^*$ is the scattering angle in the centre-of-mass frame. This radiation zero
gets diluted by the convolution with the parton densities
and by off-shell effects, but
it is responsible for the suppression of
the Born level $W\gamma$ cross section with respect to $Z\gamma$. Real radiation appearing at NLO
breaks the radiation zero,  and thus the relative impact of higher-order corrections is significantly increased.

\begin{figure}[tp]
  \centering
  \includegraphics[width=0.5\columnwidth]{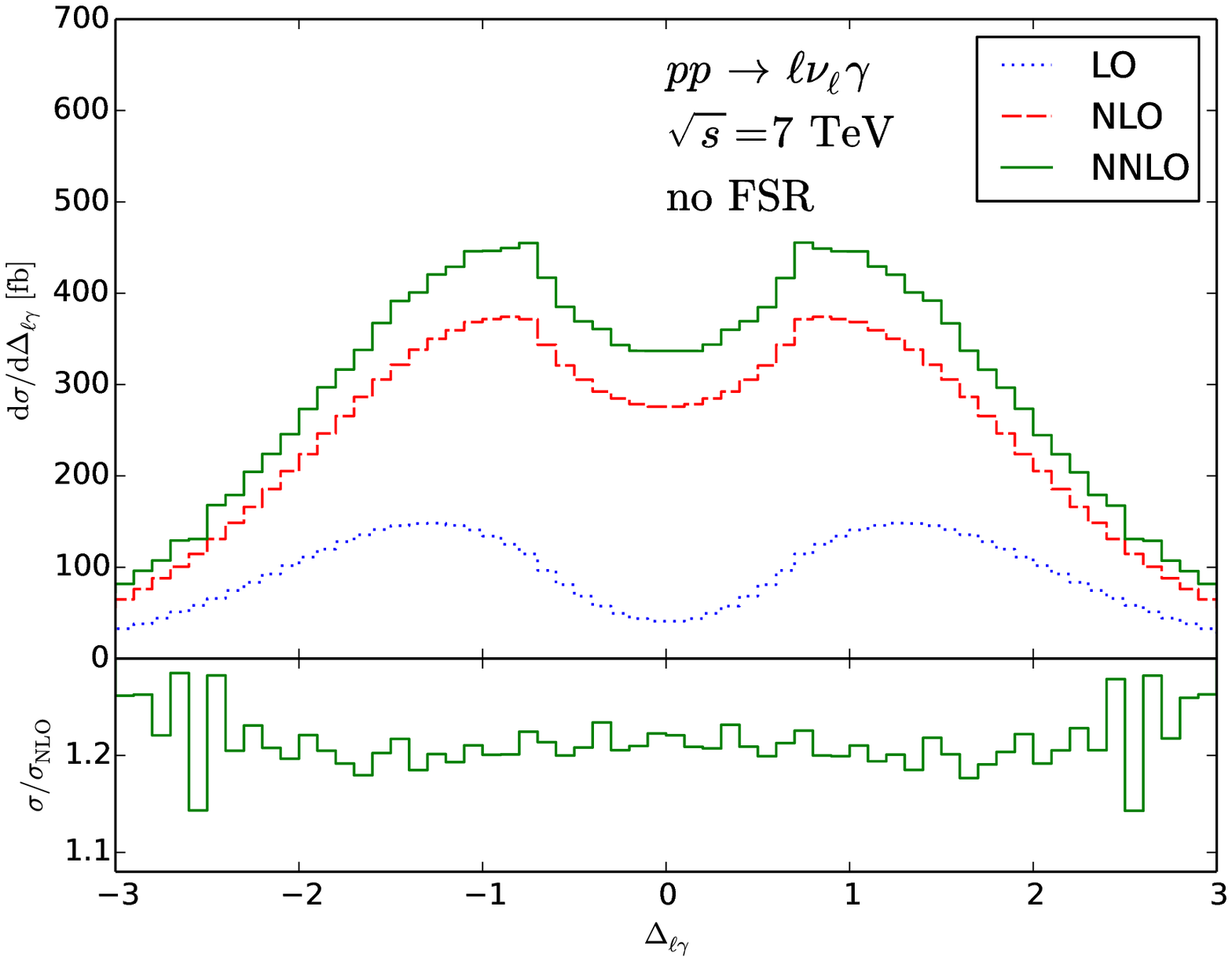}
  \hspace{-1em}
  \includegraphics[width=0.5\columnwidth]{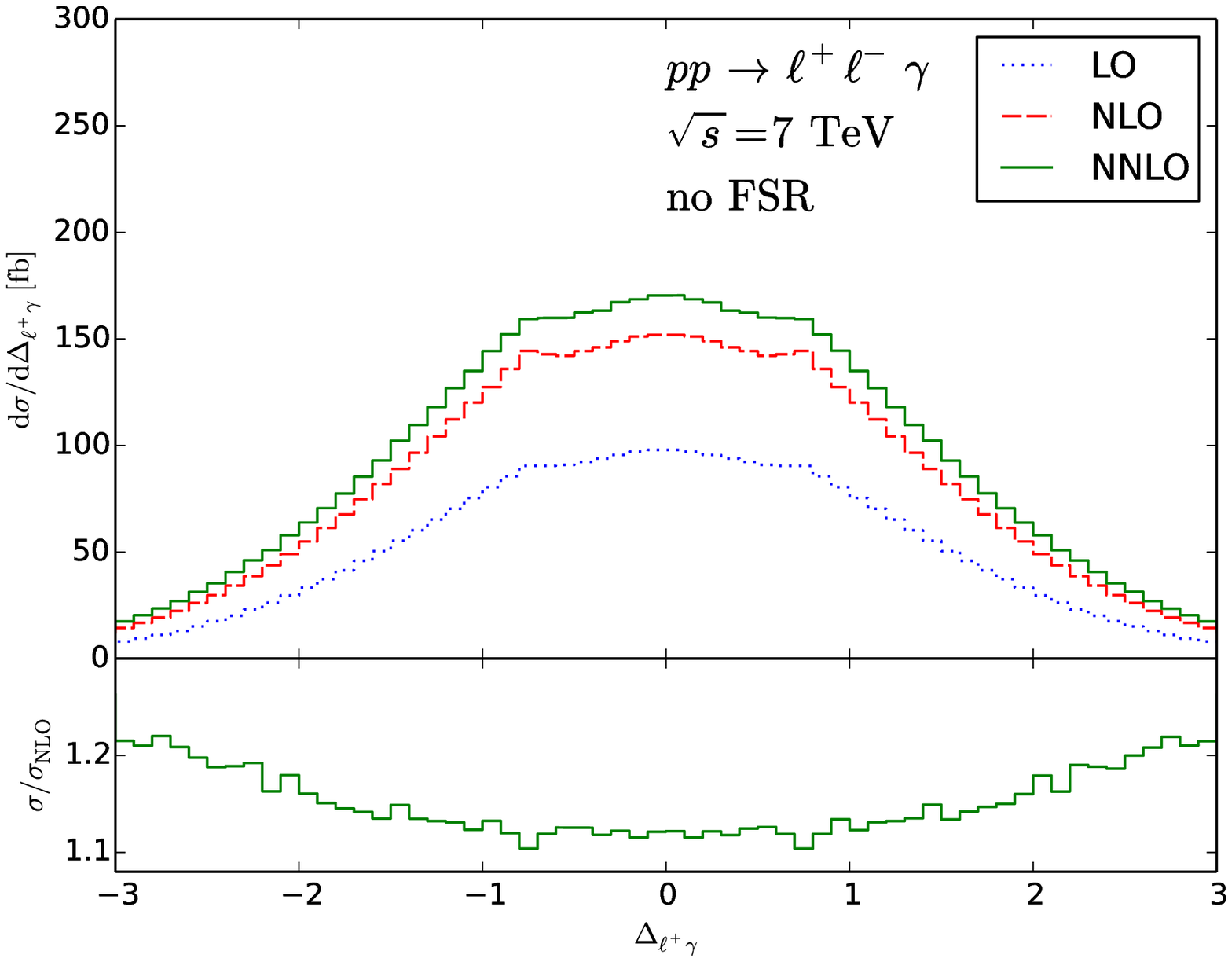}
  \caption{Rapidity difference between the charged lepton and the photon for $W\gamma$ (left) and $Z\gamma$ 
production (right) at LO (blue, dotted), NLO (red, dashed) and NNLO (green, solid). 
The lower panel shows the NNLO/NLO ratio.
Final-state radiation has been disabled for these plots.}
  \label{fig:deta}
\end{figure}

To quantitatively test this effect we consider the $pp\to \ell\nu_\ell\gamma$ and $pp\to \ell^+\ell^-\gamma$ processes
studied in \refse{sec:llgam} and \refse{sec:wgam}, with the same selection cuts.
Contrary to what was done in the previous sections we disable the contributions from final state radiation and use the narrow-width approximation.
In \reffi{fig:deta} (left) we plot the
distribution in the rapidity difference $\Delta_{\ell\gamma}$
between the charged lepton and the photon~\cite{Baur:1994sa}.

We see that the LO distribution shows a pronounced dip at central rapidities.
Although diluted by the convolution with the parton densities,
the dip is clearly visible, and is responsible for the suppression of the $W\gamma$ cross section.
Since real radiation does not respect the radiation zero, the dip is filled up by radiative corrections. 
Roughly speaking, the NLO is a {\em de facto} LO prediction in the region of the dip and the NNLO corrections are thus relatively large as well.
In contrast to $W\gamma$, the $Z\gamma$ amplitude does not exhibit a radiation zero and, consequently no dip appears in the 
rapidity-difference distribution, as can be seen in \reffi{fig:deta} (right).

The presence of the radiation zero, and of the corresponding dip in the $\Delta_{\ell\gamma}$ distribution, are thus the reason for the increased importance of radiative corrections to the $W\gamma$ processes.

\section{Summary and discussion}
\label{sec:summary}

In this paper we have presented the first complete and fully differential computation of QCD radiative corrections to $W\gamma$ and $Z\gamma$ production at hadron colliders. More precisely, we have considered the processes $pp\to \ell^+\ell^-\gamma$, $pp\to \nu_\ell\overline{\nu}_\ell\gamma$ and $pp\to \ell\nu_\ell \gamma$, where, in the first case, the lepton pair $\ell^+\ell^-$ is produced either by a $Z$ boson or a virtual photon. The diagrams in which the photon is radiated off the final-state charged leptons were consistently included.
We have presented quantitative predictions for fiducial cross sections for $pp$ collisions at $\sqrt{s}=7$ and $8$ TeV, and for various kinematical distributions (only at $\sqrt{s}=7$ TeV). The impact of QCD radiative corrections strongly depends on the applied cuts. In the case of $Z\gamma$, the impact of NNLO corrections is generally moderate, ranging from 8\% to 18\%. We have also shown that the loop induced gluon fusion contribution is generally small, and it accounts for less than 10\% of the full ${\cal O}(\as^2)$ correction.
In the case of $W\gamma$ production the NNLO effects are more important, and range from 19\% to 26\%. The larger impact of QCD radiative effects in the case of $W\gamma$ production is a well known consequence of a radiation zero~\cite{Mikaelian:1979nr} existing in the $W\gamma$ amplitude at Born level. This effect produces a suppression of the LO distribution in the rapidity difference between the charged lepton and the photon, and NLO and NNLO corrections are thus quite significant.
As expected, the impact of QCD radiative effects is strongly reduced when a jet veto is applied ($N_{\rm jet}=0$), being smaller than $3\%$ in the case of $Z\gamma$, and about $7\%$ in the case of $W\gamma$.

We add few comments on the remaining uncertainties affecting our NNLO results.
The uncertainties from missing higher-order contributions were estimated through scale variations, which are performed through independent variations of the renormalization and factorization scales around their central value (without constraints on their ratio).
In the inclusive case the NNLO scale uncertainties obtained in this way
are of the order of $\pm (1-2)\%$ in the case of $pp\to \ell^+\ell^-\gamma$ (see \refta{tab:zll_results}),
$\pm (2-3)\%$ in  the case of  $pp\to \nu_\ell\overline{\nu}_\ell\gamma$ (see \refta{tab:znunu_results}), and $\pm 4\%$ in the case of
$pp\to \ell\nu_\ell \gamma$ (see \refta{tab:wgam_results}).
The comparison of the NNLO predictions to what is obtained at NLO shows that the NNLO--NLO difference
is larger than the NLO scale dependence.
We thus conclude that, as usual, scale variations can give only a lower limit on the true perturbative uncertainty. However, the NNLO is the first order at which all partonic channels are accounted for, and we believe that the NNLO scale uncertainties obtained in the case $N_{\rm jet}\geq 0$ should provide the correct order of magnitude of the true uncertainty.
As discussed, the situation is different
for the case $N_{\rm jet}= 0$, in which
the scale uncertainties are even smaller.
Most likely, in this case a more conservative approach has to be adopted to obtain a realistic estimate
of the perturbative uncertainty (see e.g.~\cite{Stewart:2011cf,Banfi:2012jm}).

The other source of uncertainty affecting our perturbative QCD calculations is the one coming from the PDFs.
The PDF uncertainties at 68\% CL that we obtain on our fiducial cross sections are at the $1\%-2\%$ level, both at NLO and NNLO, and are thus of the same order, or smaller, than the perturbative uncertainties.
We have checked that, by using CT10~\cite{Lai:2010vv} and NNPDFs~\cite{Ball:2014uwa}, the differences we obtain with the default MMHT result
are of the same order.

The quantitative predictions we have presented for $\sqrt{s}=7$ TeV were obtained by using the same cuts
adopted by the ATLAS collaboration in their measurement of the $W\gamma$ and $Z\gamma$ cross sections~\cite{Aad:2013izg}.
We have presented a comparison to ATLAS data, both for the fiducial cross sections and for some kinematical distributions. In the case of $Z\gamma$ production the NNLO corrections slightly improve the agreement with
the data, which, however, have large uncertainties. The only exception is the case $pp\to \nu_\ell\overline{\nu}_\ell\gamma$
with a jet veto ($N_{\rm jet}= 0$), for which a $1.7\sigma$ discrepancy with the ATLAS result remains.
In the case of $W\gamma$, 
the ATLAS result overshoots the NLO prediction by
about $2\sigma$. The large NNLO corrections reduce this excess to below $1\sigma$.
The NNLO corrections improve the agreement with the data also for the kinematical distributions we have studied,
in particular for the $\pT^\gamma$ distribution. However, it is known that this distribution, at large $\pT^\gamma$, is affected by sizeable effects from EW corrections~\cite{Hollik:2004tm,Accomando:2005ra,Denner:2014bna}.
The impact of EW corrections depends on the way the additional photon radiation is treated.
It is thus difficult at present to draw conclusions on the data/theory agreement in the high-$\pT^\gamma$ region.
More generally, we think it will be important to consistently combine the QCD calculations presented in this paper with a NLO computation of EW correction, which is, however, left for future work.

\noindent {\bf Acknowledgements.}
We are grateful to Alessandro Torre for his contribution at early stages of this work.
We would like to thank Al Goshaw and Andrea Bocci for useful correspondence on the ATLAS results,
and Stefano Catani for helpful discussions. One of us (SK) would like to thank
the University of Zurich for the hospitality during the completion of this paper.
This research was supported in part by the Swiss National Science Foundation (SNF) 
under contracts CRSII2-141847, 200021-144352, 200021-156585 and by the Research Executive Agency (REA) of the 
European Union under the Grant Agreement number PITN-GA-2012-316704 ({\it Higgstools}).

\end{document}